\documentclass{article}

\usepackage{PRIMEarxiv}

\usepackage[utf8]{inputenc} 
\usepackage[T1]{fontenc}    
\usepackage{url}            
\usepackage{booktabs}       
\usepackage{amsfonts}       
\usepackage{nicefrac}       
\usepackage{microtype}      
\usepackage{lipsum}
\usepackage{fancyhdr}       
\usepackage{graphicx}       
\graphicspath{{media/}}     
\usepackage{amsmath}

\usepackage{moreverb,url}

\usepackage{xcolor}
\usepackage[colorlinks,
            bookmarksopen,
            bookmarksnumbered,
            allcolors=blue]{hyperref}

\usepackage{enumitem}
\usepackage{adjustbox}
\usepackage{graphicx}
\usepackage{subfig}
\usepackage{listings}

\newcommand{\marbl}{MARBL}

\definecolor{deepgray}{rgb}{.3,.3,.3}
\definecolor{deepblue}{rgb}{0,0,0.5}
\definecolor{deepred}{rgb}{0.6,0,0}
\definecolor{deepgreen}{rgb}{0,0.5,0}

\lstnewenvironment{cpplisting}[1][]
  {
   \ifthenelse{ \equal{#1}{} }
    {
    \lstset{
        language=C++,
        frame=tb,
        basicstyle=\footnotesize\ttfamily,
        keywordstyle=\color{deepblue}\bfseries\ttfamily,
        stringstyle=\color{deepred}\ttfamily,
        commentstyle=\color{deepgreen}\itshape\ttfamily,
        morecomment=[l][\color{magenta}]{\#},
        showstringspaces=false,
        breaklines=true,
        emphstyle=\bfseries\ttfamily
    }}
    {
    \lstset{
        language=C++,
        frame=tb,
        basicstyle=\footnotesize\ttfamily,
        keywordstyle=\color{deepblue}\bfseries\ttfamily,
        stringstyle=\color{deepred}\ttfamily,
        commentstyle=\color{deepgreen}\itshape\ttfamily,
        morecomment=[l][\color{magenta}]{\#},
        showstringspaces=false,
        breaklines=true,
        emphstyle=\bfseries\ttfamily,
        #1
    }}
  }
  {}

\title{Matrix-free approaches for GPU acceleration of a high-order finite element hydrodynamics application using MFEM, Umpire, and RAJA
}

\author{ Arturo Vargas$^\dagger$, Thomas M. Stitt$^\dagger$, Kenneth Weiss$^\dagger$, Vladimir Z. Tomov$^*$,  \\
  \textbf{Jean-Sylvain Camier$^*$, Tzanio Kolev$^*$, Robert N. Rieben$^\dagger$}, \\
  \\
  Weapons and Complex Integration, Lawrence Livermore National Laboratory, Livermore, CA, USA$^\dagger$, \\Center for Applied Scientific Computing, Lawrence Livermore National Laboratory, Livermore, CA, USA$^*$
}

\begin{document}
\maketitle

\begin{abstract}
With the introduction of advanced heterogeneous computing architectures based on GPU accelerators, large-scale
production codes have had to rethink their numerical algorithms and incorporate new
programming models and memory management strategies in order to run efficiently
on the latest supercomputers. In this work we discuss our co-design strategy to
address these challenges and achieve performance and portability with \marbl, a
next-generation multi-physics code in development at Lawrence Livermore National
Laboratory. We present a two-fold approach, wherein new hardware is used to
motivate both new algorithms and new abstraction layers, resulting in a single
source application code suitable for a variety of platforms. Focusing on \marbl's
ALE hydrodynamics package, we demonstrate scalability on different platforms and
highlight that many of our innovations have been contributed back to open-source
software libraries, such as MFEM (finite element algorithms) and RAJA (kernel
abstractions).\end{abstract}

\keywords{High-order, Finite Elements, GPUs, Performance Portability}

\section{Introduction}
The computational landscape in high-performance computing (HPC) has undergone a
dramatic change in recent years. General-purpose CPUs with large caches and
sophisticated pipeline executions have been largely replaced by low-power
data-parallel GPU accelerators in the vast majority of leadership class US
supercomputers. While this hardware shift has paved the way for large-scale
simulation codes towards exascale computing~\cite{ceed}, it has come at a
significant development cost, since HPC applications have had to rethink their
numerical algorithms and incorporate new programming models and memory
management strategies in order to run efficiently on these advanced architectures.

In this paper we discuss our experience with the GPU port of one such HPC
application, the \marbl\ code, and review our co-design strategy to address the
challenges of performance and portability across different computing architectures.
We present a two-fold approach,
wherein new hardware is used to motivate both new algorithms and abstraction
layers, resulting in a single source application code suitable for a variety of
platforms. Our work includes the development of GPU-oriented matrix-free
algorithms, porting and optimization via mini-apps, and the utilization of
open-source software libraries, such as MFEM (finite element algorithms)
\cite{anderson2021mfem}, RAJA (kernel abstraction) \cite{beckingsale2019raja} and Umpire
(memory management) \cite{beckingsale2019umpire}. We demonstrate scalability on
different platforms and highlight that many of our innovations have been
contributed back to the MFEM and RAJA libraries.

Our focus in this work is on \marbl, a next-generation multi-physics simulation
code developed at Lawrence Livermore National Laboratory for simulating high
energy density physics (HEDP) and focused experiments driven by high-explosive,
magnetic or laser based energy sources for applications including pulsed power
and inertial confinement fusion (ICF)~\cite{marbl}. \marbl\ is built on modular
physics and computer science components and makes extensive use of high-order finite element
numerical methods. Compared to standard low-order finite volume schemes,
high-order numerical methods have more resolution/accuracy per unknown and have
higher FLOP/byte ratios meaning that more floating-point operations are
performed for each piece of data retrieved from memory. This leads to improved
strong parallel scalability, better throughput on GPU platforms and increased
computational efficiency.

\begin{figure*}[tbp]
  \includegraphics[width=\textwidth]{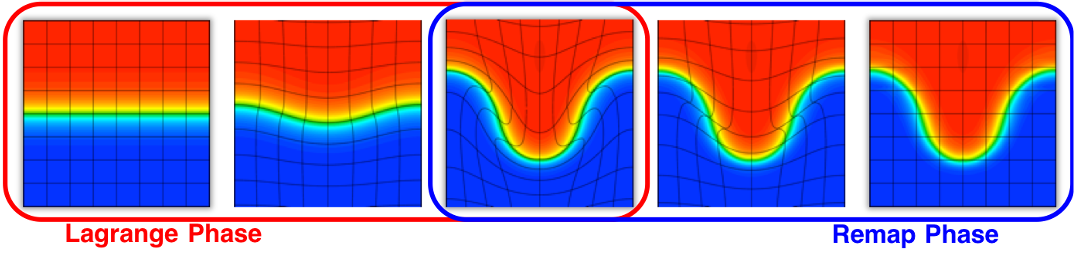}
  \caption{\marbl\ follows a three-phase high-order ALE algorithm: Lagrange, Mesh Optimization, Remap.}
\label{fig:Ale_phases}
\end{figure*}

Prior to the work described in this paper, \marbl's Arbitrary Lagrangian-Eulerian
(ALE) multi-material compressible hydrodynamics capability was an MPI-only
massively parallel code based on full matrix assembly. In the remaining sections,
we describe the process of restructuring \marbl's ALE component to an MPI + X code,
where X is a shared memory
threading model. In addition, we discuss the algorithmic refactoring which was
done to convert the full matrix assembly approach into a more efficient
and high-performance version based on ``matrix-free'' partial assembly techniques.
A key detail in our success was the co-design effort between
the application and tool libraries (MFEM, RAJA and Umpire) to steer future
directions in library developments. One of the main lessons learned in this
effort was that customization of established portability tools were crucial to
developer productivity and application performance.

The rest of this article is organized as follows:
We begin with an overview of
the high-order finite element ALE hydrodynamics approach in \marbl, and describe
key concepts of the finite element formulation and algorithmic ideas in Section~\ref{sec:overview}.
We then discuss the approach for achieving performance and portability in
Sections~\ref{sec:mfpa} and~\ref{sec:portability}.
Our two-pronged approach includes algorithmic improvements, in
particular matrix-free partial assembly, as well as performance improvements
achieved by exposing key programming features through our portable kernel layer
abstraction and memory-based optimizations through the Umpire library.
We demonstrate the overall performance and scalability of \marbl\ across both GPU and CPU platforms in
Section~\ref{sec:scaling_studies}, review our mini-app driven development in Section~\ref{sec_miniapp},
and showcase these developments in the context of high-order 3D multi-material simulations in Section~\ref{sec:high_perf_apps}.
Finally, we provide some concluding remarks in Section~\ref{sec:conclusion}.

\section{Overview of ALE}
\label{sec:overview}
\marbl\ is a multi-physics code including three temperature radiation-magneto-hydrodynamics with thermonuclear burn
for fusion plasmas. Central to it is the multi-material Arbitrary Lagrangian Eulerian (ALE) compressible hydrodynamics capability  (based on the BLAST code, see for example \cite{anderson2018high}) which we will
focus on in this paper. The high-order finite element ALE approach of \marbl\ splits the computation into three phases as illustrated in Figure~\ref{fig:Ale_phases}. The first phase (Lagrange phase) evolves the simulation within a Lagrangian frame of reference until the mesh reaches a level of distortion, the mesh is then relaxed (mesh optimization phase), and the solution is then remapped onto the optimized mesh (remap phase).
We provide a high-level overview of each of the phases in this section
and review key algorithmic implementation details in Section~\ref{sec:mfpa}.

\subsection{High-Order Lagrange Phase}
The Lagrange phase solves the time dependent Euler equations of compressible hydrodynamics within a Lagrangian frame of reference \cite{dobrev2012high}. In the case of multi-materials, $k=1, \cdots N_\text{Materials}$, we solve the following conservation laws from hydrodynamics:
\begin{align}
\rho \frac{d v}{d t} & = \nabla \cdot \sum_k \eta_k \sigma_k \label{lagEqStart} \\
\frac{d \eta_k}{d t} &= \alpha_k \\
\frac{ d \left(\eta_k \rho_k \right)}{d t} &= - \eta_k \rho_k \nabla \cdot v \\
\eta_k \rho_k \frac{d e_k}{d t} &= \eta_k \sigma_k : \nabla v - \bar{p} \alpha_k \\
\frac{d x}{d t} &= v \label{lagEqEnd}
\end{align}

Here $\rho_k$ corresponds to material density,
$v$ is the common material velocity,
$\eta_k$ serves as an indicator function which keeps track of material volume
fraction per zone, $\alpha_k$ is the rate of change in time of $\eta_k$,
$e_k$ corresponds to specific internal material energy,
$\bar{p}$ is a common thermodynamic pressure, see \cite{dobrev2016closure}, and
$x$ is the mesh node location.
The formulation makes use of the material derivative
\[
\frac{d \alpha}{d t} = \frac{\partial \alpha}{\partial t} + v \nabla \cdot \alpha,
\]
as well as a total stress tensor for each material, $\sigma_k$, which is the sum of the physical stresses and stress due to artificial viscosity.

Building from the MFEM finite element library, \marbl\ uses arbitrary high-order finite elements to solve Equations (\ref{lagEqStart})--(\ref{lagEqEnd}). A continuous Galerkin discretization is used for the kinematic variables (mesh velocity and position), while a discontinuous Galerkin discretization is used for the thermodynamic variables (indicator functions, density, and energy). The approach presented here can be viewed as a generalization of the traditional staggered-grid discretization for hydrodynamics (SGH)~\cite{barlow2016arbitrary}.

The finite element algorithm to solve Equations (\ref{lagEqStart})--(\ref{lagEqEnd}) requires solving a global linear system for the momentum conservation equation. The key operators in this algorithm are the ``mass'' matrix and a ``force'' matrix connecting the kinematic and thermodynamic variables. Computation of a hyper-viscosity term for shock capturing  requires an additional
``stiffness matrix'' operator as described in \cite{maldonado2020}. The traditional approach of fully assembling corresponding matrices for these operators has been found to have poor scaling characteristics motivating the development of a matrix-free approach.


\subsection{Mesh Optimization}
\label{sec_tmop}

The mesh optimization algorithm in \marbl\ is based on the Target-Matrix
Optimization Paradigm (TMOP)~\cite{dobrev2019target, knupp2012introducing}
for high-order meshes.
The method is based on node movement and does not alter the mesh topology.
Optimal mesh node positions are determined by minimizing a nonlinear
variational functional.
One of the main advantages of this approach is that all computations can be
expressed through finite element operations, making the algorithm well-suited for
matrix-free partial assembly implementation as discussed in Section~\ref{sec_tmop_pa}.
The TMOP objective function has the form:
\begin{equation}
\label{eq_tmop_F}
  F(x) = \int_{\Omega} \mu(T(x)) +
         \gamma \int_{\Omega} \left(\frac{x - x_0}{d(x_0)}\right)^2,
\end{equation}
where $x$ is the vector of mesh positions, $T(x)$ is a Jacobian matrix that
represents the transformation between user-specified target elements and
physical elements, and $\mu$ is a nonlinear function that measures mesh quality.
The second term in \eqref{eq_tmop_F} is used to limit the node displacements
based on the maximum allowed displacements $d(x_0)$ that are defined with
respect to the initial mesh position $x_0$.
The normalization constant $\gamma$ balances the magnitudes of the two terms;
its value is computed once on the initial mesh.

The objective $F(x)$ is minimized by solving the nonlinear problem
$\partial F(x) / \partial x = 0$ with the Newton's method.
This requires the ability to compute the gradient $\partial F(x) / \partial x$
and to apply/invert the Hessian $\partial^2 F(x) / \partial x^2$ of the objective
functional.
In addition, TMOP has the ability to perform simulation-based
mesh adaptivity by querying the values of discrete function $\xi$ provided
by the simulation \cite{dobrev2020adaptivity}. For example, $\xi$ can prescribe
local mesh size based on the current state of the physical system.
Since $\xi$ is only defined with respect to the initial mesh, the adaptivity
process includes a procedure to solve a pseudo-time advection equation, namely,
\begin{equation}
\label{eq_tmop_xi}
  \frac{d \xi}{d \tau} = v \cdot \nabla \xi,
\end{equation}
which is a method to advect the function $\xi$ between different meshes, as done
in the remap phase of the code, see Section~\ref{sec:Remap}.
All methods described in this section are based on standard finite element
operations that can be performed in matrix-free manner.

\subsection{Remap Phase}
\label{sec:Remap}
Following the mesh optimization, the ALE remap phase is responsible for transferring field quantities from the current Lagrangian mesh to the optimized mesh.
The transfer process is posed as the advection of field values between the two meshes over a fictitious time interval, $\Delta \tau$. For each material $k$,
we form and solve the following set of advection equations:
\begin{align}
\frac{d \left( \rho v \right)}{d \tau} &= u \cdot \nabla \left(\rho v \right) \label{eq:momentum-remap}\\
\frac{d \textcolor{black}{\eta_k}}{d \tau} &= u \cdot \nabla \textcolor{black}{\eta_k} \\
\frac{d \textcolor{black}{\langle \eta \rho \rangle_k}}{d \tau} &= u \cdot \nabla \textcolor{black}{\langle \eta \rho \rangle_k} \\
\frac{d \textcolor{black}{\langle \eta \rho e \rangle_k}}{d \tau} &= u \cdot \nabla \textcolor{black}{\langle \eta \rho e \rangle_k} \\
\frac{d x}{d \tau} &= u.
\end{align}
Here $u$ corresponds to the mesh displacement, and the field variables are the same as those in the Lagrange phase Equation(\ref{lagEqStart})--(\ref{lagEqEnd}).

For the transfer of the momentum field \eqref{eq:momentum-remap} we follow a standard continuous Galerkin discretization. Since only the action of the operators are needed, we bypass the assembly of the global matrices.  For the transfer of the remaining fields, it is imperative that the transfer procedure is conservative, accurate, and is bounds preserving, i.e. the solution remains in a set of admissible bounds.  In our target application of remapping multi-material fields, preservation of solution bounds plays a critical role at material interfaces in order to achieve high-fidelity simulations. For ALE remap we use the approach described in \cite{anderson2015monotonicity}. The underlying strategy is to begin with a low-order solution which is guaranteed to be within bounds, and use a separate high-order solution to make corrections to the low order solution. The correction strategy draws on principles from the flux corrected transport (FCT) work \cite{kuzmin2012flux,anderson2015monotonicity}. These nonlinear corrections require algebraic modifications to the fully assembled advection matrix in the DG-FCT scheme, so the current implementation cannot be done matrix-free. However, we are actively researching and developing
a new DG remap method which retains all of the numerical benefits but can be done in a high-performance matrix-free manner.

\section{Matrix-Free Partial Assembly}
\label{sec:mfpa}
For many applications, the assembly of large global sparse matrices can be a performance bottleneck. Computation with the common CSR matrix format also leads to indirection access, which further limits performance. In most cases, the assembly of the finite element operators in the ALE phases can be avoided as it is only the action of the operator that is needed. We leverage this by using matrix-free algorithms based on the partial-assembly approach described in~\cite{anderson2021mfem} and available in the MFEM library. Under partial assembly, FEM operators are decomposed into a sequence of operators with cascading scope only requiring quadrature-specific data. For completeness we provide a brief overview of the approach; we refer the reader to~\cite{anderson2021mfem} for additional details and to the MFEM library~\cite{mfem-web} for open-source implementation.

\begin{figure*}[tbp]
  \centering
  \includegraphics[width=.8\textwidth]{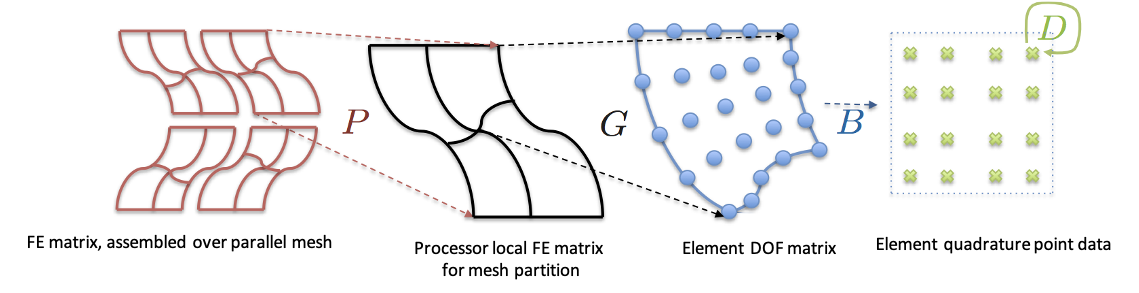}
  \caption{Finite element operator decomposition from global operator to element quadrature data.}
  \label{fig:OpDecomp}
\end{figure*}

\subsection{Operator Decomposition}
\label{sec:opDec}
Following the nomenclature introduced in \cite{anderson2021mfem}, for a given global finite element operator ($A$) distributed over a number of processors, we recognize that its assembly or evaluation can be decomposed into a product of operators with cascading scope corresponding to parallel ($P$), mesh ($G$), basis ($B$), and quadrature-specific parts ($D$).
\[
\boxed{A = P^{T} G^{T}B^{T} D B G P}
\]

Starting with a distributed mesh (as depicted in Figure~\ref{fig:OpDecomp}),
the role of the $P/P^{T}$ operators is to perform the transfer from a global to a processor local domain.
The operators $G/G^{T}$ perform an additional restriction/prolongation at the element level, while $B/B^T$ hold the values of the basis functions sampled
at quadrature points. For simplicity we assume that test and basis functions are the same, but note that this is not required. Lastly, $D$ corresponds to element-specific quadrature data. Under the partial-assembly methodology each of these operators are precomputed and stored.

\subsection{Accelerated Matrix-Vector Evaluation Using Sum-Factorization}
\label{sec_factorization}
Building on the concepts of operator decomposition, we restrict our discussion to quad and hexahedral elements thereby enabling us to exploit the tensor product structure of the basis functions. Continuing with the nomenclature introduced in Section~\ref{sec:opDec}, we focus on the element local matrix-vector product,
\begin{equation}
y = B^{T} D B x. \label{eq:local_action}
\end{equation}
Using the element level mass matrix, $M_E$, to lead our discussion we express the construction as
\[
(M_E)_{i,j} = \sum_{k=1}^{nq} \alpha_k \phi_j(q_k) \phi_i(q_k) |J_E(q_k)| \quad i,j=1,...,nq.
\]
Here $\alpha_k$ corresponds to quadrature weights, $\phi_i$ are the basis functions sampled at quadrature points, $q_k$, and $|J_E(q_k)|$ is the geometric factor between reference to physical element. When using quad and hex elements, the multi-dimensional basis functions are simply tensor products of their one-dimensional counterparts. In 2D the basis functions have the form
\[
\phi_i \left(q_k \right) = \phi_{i_1}^{1d} \left(q_{k_1} \right) \times \phi_{i_2}^{1d} \left(q_{k_2} \right).
\]
Recognizing this decomposition enables us to decompose the operator $B$ in Equation\ref{eq:local_action} into
\[
B = B^{2d}_{k1,k2,i1,i2} = B^{1d}_{k1, i1} B^{1d}_{k2,i1}  =   \phi_{i_1}^{1d} \left(q_{k_1} \right) \times \phi_{i_2}^{1d} \left(q_{k_2} \right),
\]
where $B^{1d}$ are the one-dimensional basis functions. Finally, by arranging $D$ as a two dimensional array, computing the action of $M_E$ may be expressed as a sequence of tensor contractions
\[
\begin{split}
y &= (M_E)_{i_1,i_2,k_1,k_2} x_{i_1,i_2} \\
  &= B^{1d}_{k_1,i_1} B^{1d}_{k_2,i_2} \left(D_E \right)_{k_1,k_2}
     B^{1d}_{k_1,j_1} B^{1d}_{k_2,j_2}  x_{j_1,j_2}.
\end{split}
\]
These tensor contractions can be implemented as a sequence of dense matrix-matrix
multiplications suitable for single instruction multi-data (SIMD) parallelism.

To illustrate the advantages of the partial assembly approach, Table~\ref{table:FAvsPA} compares storage and algorithmic complexity between full assembly and partial assembly of matrix-vector evaluations per element.
As an example, when $d=3$ and $p=3$, the storage for 1000 elements would require around 729K floating point values, while the partial assembly approach would only require storing 27K floating point values; similar trends follow when considering floating-point operations for assembly and applying the action of the operator.
In summary, partial assembly provides clear advantages in terms of both memory requirements and floating-point operations.

\newpage
\begin{center}
\begin{table}[h!]
\centering
  \caption{\small{Comparison of storage and assembly/evaluation FLOPs required for full and
  partial assembly algorithms on tensor-product element meshes (quadrilaterals
  and hexahedra). Here, \(p\) represents the polynomial order of the basis
  functions and \(d\) represents the number of spatial dimensions. The number
  of DOFs on each element is \(\mathcal{O}(p^{d})\) so the ``sum factorization
  full assembly'' and ``partial assembly'' algorithms are nearly optimal.
  }}
  \label{table:FAvsPA}
  \begin{tabular}{l@{\hspace{0.5em}}lll}
    \toprule
    \textbf{Method} & \textbf{Storage} & \textbf{Assembly} & \textbf{Evaluation} \\
    \midrule
    Traditional full assembly + matvec & \(\mathcal{O}(p^{2d})\) & \(\mathcal{O}(p^{3d})\) & \(\mathcal{O}(p^{2d})\) \\
    Sum factorized full assembly + matvec & \(\mathcal{O}(p^{2d})\) & \(\mathcal{O}(p^{2d+1})\) & \(\mathcal{O}(p^{2d})\) \\
    Partial assembly + matrix-free action  & \(\mathcal{O}(p^{d})\) & \(\mathcal{O}(p^d)\) & \(\mathcal{O}(p^{d+1})\)\\
  \bottomrule
  \end{tabular}
\end{table}
\end{center}

\subsection{Partial Assembly in Lagrange Phase}
Applying the finite element discretization to Equations (\ref{lagEqStart})--(\ref{lagEqEnd}) leads to the so-called ``Semi-discrete'' set of equations, which under the appropriate initial conditions and boundary conditions can be solved to a final time with numerical integration techniques.
 \begin{align}
M_v \frac{d v}{dt} &= - \sum_k F_k 1 \label{eq:semiDiscreteLagStart} \\
\frac{d \eta_k}{d t} &= \alpha_k \\
\eta_k \rho_k |J| &= \eta_{k_0}\rho_{k_0}|J_0| \\
M_{\mathcal{E},k} \frac{d e_k}{d t} &= F^{T}_k v - c_k \\
\frac{d x}{d t} &= v \label{eq:semiDiscreteLagEnd}
\end{align}

The key operators in equations (\ref{eq:semiDiscreteLagStart})--(\ref{eq:semiDiscreteLagEnd}) are the Mass matrix, $M_{v}$, arising from the conservation of momentum in (\ref{eq:semiDiscreteLagStart}), and the force matrix, $F_k$, connecting the kinematic and thermodynamic spaces. There is an additional operator required for computing the hyper-viscosity operator as discussed in \cite{maldonado2020}. Both $M_v$ and $F_v$ are global operators and applying their action requires global communication in a distributed memory setting, for example using domain decomposition with MPI. Based on the memory and algorithmic complexity requirements as presented in Table~\ref{table:FAvsPA}, our algorithms bypass the full matrix-assembly, and rely on the partial-assembly approach when applying the action of these operators.


\subsection{Matrix-Free Mesh Optimization Phase}
\label{sec_tmop_pa}
The computations in this phase utilize a fully matrix-free approach, i.e., all data
that is needed at quadrature points is computed on-the-fly and there is no matrix $D$.
This decision is motivated by the complex structure of the Hessian of $F(x)$ and the
fact that stored quadrature data cannot be reused
in consecutive calculations.
The major computational kernels in this phase are related to the first and
second derivatives of the $\mu$-integral in \eqref{eq_tmop_F}.
These are given here to provide the relevant context for the discussion:

\begin{equation}
\begin{split}
\label{eq_tmop_F1}
\frac{\partial F_{\mu}}{\partial x_{a,i}} =
  & \sum_q^{N_q} w_q \det(W) \\
    & \sum_{m,n = 1}^d \frac{\partial \mu}{\partial T_{mn}}
      \sum_{l=1}^d \delta_{m, a} \frac{\partial w_i}{\partial \bar{x}_l} (W^{-1})_{ln},
\end{split}
\end{equation}
\begin{equation}
\label{eq_tmop_F2}
\begin{split}
& \frac{\partial^2 F_{\mu}}{\partial x_{b,j} \partial x_{a,i}}
  =
  \sum_q^{N_q} w_q \det(W) \hspace{-1mm}
    \sum_{m,n = 1}^d \sum_{o,p = 1}^d
      \bigg[ \frac{\partial^2 \mu}{\partial T_{op} \partial T_{mn}}  \\
  & ~
      \left( \sum_{l=1}^d \delta_{o, b} \frac{\partial w_j}{\partial \bar{x}_l} (W^{-1})_{ln} \right)
      \left(\sum_{l=1}^d \delta_{m, a} \frac{\partial w_i}{\partial \bar{x}_l} (W^{-1})_{ln} \right)
      \bigg].
\end{split}
\end{equation}
The above derivatives are taken with respect to the mesh position vector
$x = (x_1 \dots x_d)^T$ where $d$ is the dimension and each component's
expansion is $x_a = \sum_i x_{a,i} w_i(\bar{x}), a = 1 \dots d$.
These integral computations use a standard Gauss-Legendre quadrature rule with
$N_q$ points, $w_q$ are the corresponding quadrature weights,
$W$ is a $d \times d$ matrix, and $\delta$ is the Kronecker delta.

In addition to the standard coefficients and geometric data that is required to
compute the integrals, the quadrature-based GPU calculations include the following:
the target matrices $T(x)$ and the mesh quality metric $\mu(T)$;
the first derivatives $\partial \mu / \partial T$, which are $d \times d$
matrices; the second derivatives $\partial^2 \mu / \partial T^2$, which are
4-tensors of dimension $d \times d \times d \times d$.

Matrix-free GPU kernels were developed for the computation of the objective
\eqref{eq_tmop_F}, the assembly of the nonlinear residual vector \eqref{eq_tmop_F1},
and the action of the Hessian \eqref{eq_tmop_F2}.
Furthermore, diagonal Jacobi preconditioning of the Hessian operator
$\partial^2 F / \partial x^2$ is performed by a matrix-free computation of the
diagonal of \eqref{eq_tmop_F2}.
All of the above kernels utilize the tensor structure of the basis
functions $w_i$ to perform efficient tensor contractions and
achieve the optimal operation complexity, see Section~\ref{sec_factorization}.

GPU support is also provided for TMOP's mesh adaptivity capabilities.
In this case, the construction of the matrix $T(x)$ depends on a discrete
function $\xi$ that must be advected between different meshes through
\eqref{eq_tmop_xi}. As \eqref{eq_tmop_xi} is a standard advection equation, we
make direct calls to MFEM's existing internal partial assembly kernels for the
action of the mass and convection operators.

\subsection{Partial Assembly in Remap Phase}
Lastly, for the Remap phase the set of semi-discrete equations take the form of
\begin{align}
M_v \frac{d v}{d \tau} &= K_v v \label {eq:semiDiscreteRemapStart} \\
M^{*} \frac{d \eta_k}{d \tau} &= K^* \eta_k \\
M^* \frac{d \langle \eta \rho \rangle_k }{d \tau} &= K^* \langle \eta \rho \rangle_k \\
M^* \frac{d \langle \eta \rho e \rangle_k }{d \tau} &= K^* \langle \eta \rho e \rangle_k \\
\frac{d x}{d \tau} &= x_{\tau=1} - x_{\tau=0}
\end{align}
where  $M_{v}$ and $K_{v}$ and the ``mass'' and ``convection'' matrices corresponding to the continuous Galerkin discretization. As only the action is required, we follow the operator decomposition in Section~\ref{sec:opDec} to bypass matrix assembly.
The remaining operators, $M^{*}$ and $K^{*}$, are algebraically modified ``mass'' and ``convection'' matrices
from the Algebraic DG-FCT discretization discussed in Section~\ref{sec:Remap}. Due to algebraic modifications, the operators currently require full matrix assembly. An active area of research is the development of a matrix-free remap framework for \marbl\ .

\section{Platform Portability Through MFEM, RAJA, and Umpire}
\label{sec:portability}
The effort to refactor \marbl's ALE algorithms and data structures was a
multi-year process that required a number of cross-project collaboration; it was
an iterative process that provided many opportunities for innovation and
application-driven development. In this section, we describe the integration of
technology and restructuring of \marbl\ to improve productivity and code
performance.

Since MFEM serves as a toolkit for developing algorithms within \marbl, introducing GPU
capabilities into MFEM was a natural starting point for \marbl's platform portability.
With the release of MFEM 4.0, there were a number of features that provided the foundation for
applications to begin leveraging GPUs, including fundamental finite element routines such
as computing geometric factors, quadrature interpolators, as well as matrix-free action
for a subset of its finite element operators.

MFEM also introduced intermediate layers between memory allocations and
kernel execution to support GPU acceleration.
The intermediate layer for kernels enables users to choose a backend at run-time;
supported backends include
  the RAJA abstraction layer~\cite{beckingsale2019raja},
  native OpenMP,
  CUDA,
  HIP,
  and OCCA~\cite{medina2014occa}.
Additionally, data structures in MFEM such as vectors,
arrays, and matrices are equipped with an intermediate layer, which keeps
track of host and device pointers and coordinates transfers between
different memory space.

\subsection{Development with RAJA in \marbl}
\label{raja_in_marbl}

After updating to MFEM 4.0, \marbl\ gained the
capability to perform basic memory movement between host and device
as well as the ability to offload a select number of MFEM routines to a device.
To further enable GPU capabilities, the RAJA abstraction layer was
chosen to express computational kernels within \marbl.

With the RAJA abstraction layer, loop bodies are captured in C++ lambdas and
template specialization via ``execution policies'' to select an
underlying programming model, i.e., Sequential, OpenMP, CUDA, or HIP. To compliment
MFEM's run-time selectivity, development in \marbl\ is done through a macro layer
that forwards the lambda to the RAJA method that matches the MFEM configured
backend (i.e. CPU, OpenMP, CUDA, or HIP).  Listings~\ref{C_style_for},~\ref{RAJA_style_for}
and~\ref{Marbl_style_for} compare a basic loop expressed using standard C, native RAJA, and
the \marbl-RAJA abstraction layer, respectively.

\begin{cpplisting}[caption={C-style for loop}, label={C_style_for}, float]
double* x; double* y;
double a;
double tsum = 0;
double tmin = MYMAX;

for (int i = 0; i < N; ++i) {
  y[i] += a * x[i];
  tsum += y[i];
  if (y[i] < tmin) { tmin = y[i];}
}
\end{cpplisting}

\begin{cpplisting}[caption={RAJA-style for loop},  label={RAJA_style_for}, float]
double* x; double* y;
double a;
RAJA::ReduceSum<red_pol, double> tsum(0);
RAJA::ReduceMin<red_pol, double> tmin(MYMAX);

RAJA::forall<exec_policy>(
    RAJA::RangeSegment(0, N),
    [=] (int i) {
  y[i] += a*x[i];
  tsum += y[i];
  tmin.min(y[i]);
});
\end{cpplisting}

\begin{cpplisting}[caption={MARBL-style for loop}, label={Marbl_style_for}, float]
double *x; double *y;
double a;
BlastReduceSum tsum(0);
BlastReduceMin tmin(MYMAX);

BLAST_FORALL(i, 0, N, {
  y[i] += a*x[i];
  tsum += y[i];
  tmin.min(y[i]);
});
\end{cpplisting}

For most of the initial refactoring in \marbl, the RAJA ``forall'' method was the work horse
for developing portable kernels. The simplicity of RAJA forall made it easy to
integrate, but the simplicity also limited the exposed parallelism, which
limited performance and motivated the search for abstractions that could better
utilize the hardware. While
more detailed threading was available through the RAJA ``kernel'' interface,
we determined that the associated per-kernel policies were too complex to maintain for our ALE hydro application.
Through a co-design effort that leveraged expertise from the RAJA, MFEM, and \marbl\ teams, we developed
the ``RAJA Teams'' API, which is available in RAJA v0.12+ as an experimental feature.

\subsection{Enhancing Performance with RAJA Teams}
To develop the RAJA Teams abstration layer, we used the following guiding principles:
\begin{description}[leftmargin=1em]
\item[Simplicity:] The API should be simple to understand and use
  so that developers can focus on algorithms.
\item[Expose key features:] For GPUs, we found that the use of
  device shared memory improved performance and an abstraction is needed so algorithms remain correct when executed on the CPU.
\item[Lightweight:] Abstraction layers should add minimal overhead
  and allow the compiler to optimize.
\end{description}

We developed the RAJA Teams framework using a bottom-up approach, in which
we built an abstraction layer around loop patterns arising from partial-assembly
based algorithms for high-order finite elements.
Listing~\ref{Marbl_style_threads} illustrates an example of the RAJA Teams API
to load quadrature data from a \texttt{DenseTensor} into a shared memory array.

The RAJA Teams API is composed of three core concepts. The first is the
\textbf{Launch Method}, which is responsible for creating a kernel execution
space.  The launch method uses a run-time value, \texttt{ExecPlace}, and a
grid configuration of teams and threads to switch between host or device execution.
The teams and threads concept is based on CUDA's and HIP's programming model
in which computation is performed on a predefined grid composed of threads, which are
then grouped into blocks (teams in RAJA).  Second, the \textbf{Launch Context}
in the launch lambda is used for control flow within the kernel, e.g. to perform
thread synchronizations as would be done with CUDA's or HIP's \texttt{\_\_syncthreads()}
routine.  Lastly, within the kernel execution space, algorithms are expressed
in terms of a nested loop hierarchy. Hierarchical parallelism is naturally
expressed through RAJA \textbf{Loop Methods}, where, in CUDA, `outer' loops are mapped to
block-based policies and inner loops may be mapped to thread policies.

In addition, because the utilization of GPU shared memory is an important
cornerstone for kernel optimizations, the RAJA teams API includes a macro,
\texttt{RAJA\_TEAM\_SHARED}, which expands out to device shared memory or host
stack memory, depending on the desired execution space.

To simplify development within \marbl, we also developed a macro layer
over the RAJA Teams API similar to our RAJA forall macro wrapper.
Listings~\ref{CUDA_style_threads},~\ref{RAJA_style_threads} and~\ref{Marbl_style_threads} compare the
native, RAJA, and MARBL kernel APIs.
As can be seen in Listing~\ref{Marbl_style_threads}, the macros allow us to remove
much of the boilerplate code that would be common to all Teams-based kernels
over mesh elements and quadrature points.

\begin{cpplisting}[caption={CUDA-style hierarchical parallelism}, label={CUDA_style_threads}, float]
DenseTensor A(D1D, Q1D, NE);

void kernel(DenseTensor A) {

 const int z = blockIdx.x;
 __shared__ double s_A[max_D1D][max_Q1D];

 // Load into shared memory
 for(int q=threadIdx.y; q<Q1D; q+=blockDim.y){
  for(int d=threadIdx.x; d<D1D; d+=blockDim.x){
   s_A[d][q] = A(d, q, z);
  }
 }

 // Synchronize threads in a thread team
 __syncthreads();
}
\end{cpplisting}

\begin{cpplisting}[caption={RAJA-style hierarchical parallelism},  label={RAJA_style_threads}, float]
using namespace RAJA;
using namespace RAJA::expt;
DenseTensor A(D1D, Q1D, NE);

RangeSegment team_range(0, NE);
RangeSegment q_range(0,Q1D);
RangeSegment d_range(0,D1D);

launch<launch_policy>(
    ExecPlace, // runtime selection
    Grid(Teams(NE), Threads(Q1D, Q1D)),
    [=] RAJA_HOST_DEVICE (LaunchContext ctx) {
  // Loop over elements
  loop<team_x>(ctx, team_range, [&](int z) {
   RAJA_TEAM_SHARED double s_A[max_D1D][max_Q1D];

   // Load into shared memory
   loop<thread_y>(ctx, q_range, [&](int q) {
    loop<thread_x>(ctx, d_range, [&](int d) {
     s_A[d][q] = A(d, q, z);
    });
   });

   // Synchronize threads in a team
   ctx.teamSync();
 });
});
\end{cpplisting}

\begin{cpplisting}[caption={MARBL-style hierarchical parallelism}, label={Marbl_style_threads}, float]
DenseTensor A(D1D, Q1D, NE);

// Loop over elements
BLAST_FORALL_TEAM(ctx, z, NE, D1D, Q1D, {
  BLAST_SHARED double s_A[max_D1D][max_Q1D];

  // Load into shared memory using 2D thread loop
  TEAM_LOOP_2D(ctx, d, q, D1D, Q1D, {
    s_A[d][q] = A(d, q, z);
  });

  TEAM_SYNC;
});
\end{cpplisting}


\subsection{Enhancing Memory Management with Umpire}
During \marbl's initial GPU refactoring effort, we quickly realized that
device memory allocations (e.g.\ device \texttt{malloc}s and device \texttt{free}s)
were drastically limiting run-time improvements; many kernels used local stack-allocated or fixed
lifetime temporary (scratch) memory that introduced expensive global
synchronizations in every time step. With application run time in mind we
decided to move these large localized allocations into permanent memory
(allocations lasting for the duration of the simulation). Finding
performance-impacting allocations was tedious as our approach relied on being
able to match up \texttt{cudaMalloc} API calls in the profiler timeline with our
in-source annotations.

Although this initial approach eliminated the runtime impacts of repeated allocations,
it increased our overall device memory usage, which impacted our ability to run larger problems.
Since our initial performant implementation
resulted in many temporary buffers being moved into permanent memory, we devised
a strategy in which we could reuse memory buffers in parts of the code where
short term scratch buffers were required.
To assist in the effort, we integrated the Umpire memory resource
library~\cite{beckingsale2019umpire}.  With Umpire, developers can create
multiple device allocators and configure them to use either the standard device
malloc or special purpose allocators like pools. With pools a large chunk of
memory is allocated upfront and subsequent allocations are pulled from the pool.
The total size of pools in Umpire can change over time, either when more memory
is needed than the pool currently has or when a coalesce operation is requested.
Pools are specifically designed to enable fast memory allocations and
deallocations and are ideal for temporary or scratch memory that may be needed
in kernels.


As previously mentioned, \marbl\ is built on MFEM and nearly all of its data
structures related to device computations are MFEM objects. With the MFEM
4.0 release all MFEM data containers are built on an internally-defined
\texttt{Memory<T>} memory buffer container. Similar to MFEM's
developer-configurable \texttt{Device} policy, the memory policies used are
configurable at run-time. 
Each \texttt{Memory} object contains a host pointer and, if a compute device is
being used, a device pointer.  These \texttt{Memory} objects know how to
allocate and copy memory between their associated spaces.
They also track the validity of the underlying data in each space (host, device, or both),
which greatly aids memory synchronization and debugging.
%
%
As part of our porting effort, we enhanced MFEM's memory manager to support Umpire allocators.
We also introduced finer-grained controls on per-allocation setup to make it easy to use a single
permanent allocator by default, but a temporary or scratch allocator when appropriate.

In order to measure memory utilization, we ran a representative 3D shaped charge
problem on 12 GPUs with ALE and TMOP. When compared to our initial performant
implementation that used 181~GB of device memory, our memory-optimized version
used only 97.3~GB, a reduction of nearly 50\%, of which 55\% (53.6~GB) is now in a
shareable temporary allocator.  This temporary allocator can be used by other
packages in \marbl\ that, when also configured to use Umpire, further reduce
total memory utilization in multi-physics/multi-package simulations.

\subsubsection{Thread Local Memory}

After we started seeing diminishing returns on managing memory spaces and
lifetimes, we began to optimize usage of our allocated memory.
The CUDA C API makes it easy to query runtime memory usage but it inform on ``how'' your memory is actually
being used. Notably, while much of the device memory is explicitly allocated by \marbl\
itself, some is allocated by the CUDA runtime, and for large problems we observed
that
our explicitly allocated device memory was only around 70\% of the total allocated device memory.
Further investigation revealed that this was due to CUDA runtime allocations
for kernels with large amounts of spilled thread-local-memory.

As discussed in Section~\ref{raja_in_marbl}, RAJA forall was used as the initial
abstraction for device-friendly kernels and while a number of kernels were
revisited and ported to RAJA Teams for improved performance, there were still
many that existed in their original forall state. A subset of our remaining
RAJA forall kernels used large iteration-local stack arrays of fixed size based
on our maximum allowable polynomial order. Our maximum  order requires much more
iteration-local memory than is needed in many simulations and, crucially, even
if a simulation is running at a higher order the iteration-local arrays cause a
CUDA runtime allocation that is never released; if there is just one kernel that
has large thread-local storage requirements, those requirements will effect the
memory availability for the entire simulation.
Specifically, the CUDA runtime will allocate
\texttt{(bytes per thread) $\cdot$ (max threads per SM) $\cdot$ (number of SMs)} bytes.

To find the kernels that were causing noticeable increases in device memory usage
we intercepted calls to \texttt{cudaLaunchKernel} and measured the device memory
availability before and after. We then incrementally refactored these
kernels to a teams-style shared memory implementation or, for kernels where this
wasn't appropriate, we pre-allocated the thread-local memory in our Umpire
temporary pool. Pre-allocating global memory resulted in a roughly 2x slowdown for
each kernel modified this way but, for \marbl\, this was acceptable because the
kernels in this group accounted for a negligible amount of runtime both before
and after the refactor.

This refactoring removed the use of thread-local arrays owned by the CUDA runtime
and freed up more than 2~GB of additional memory per device,
a significant improvement considering that our GPUs have only 16~GB of memory.

\begin{figure}[tbp]
\centering
  \includegraphics[scale=0.35]{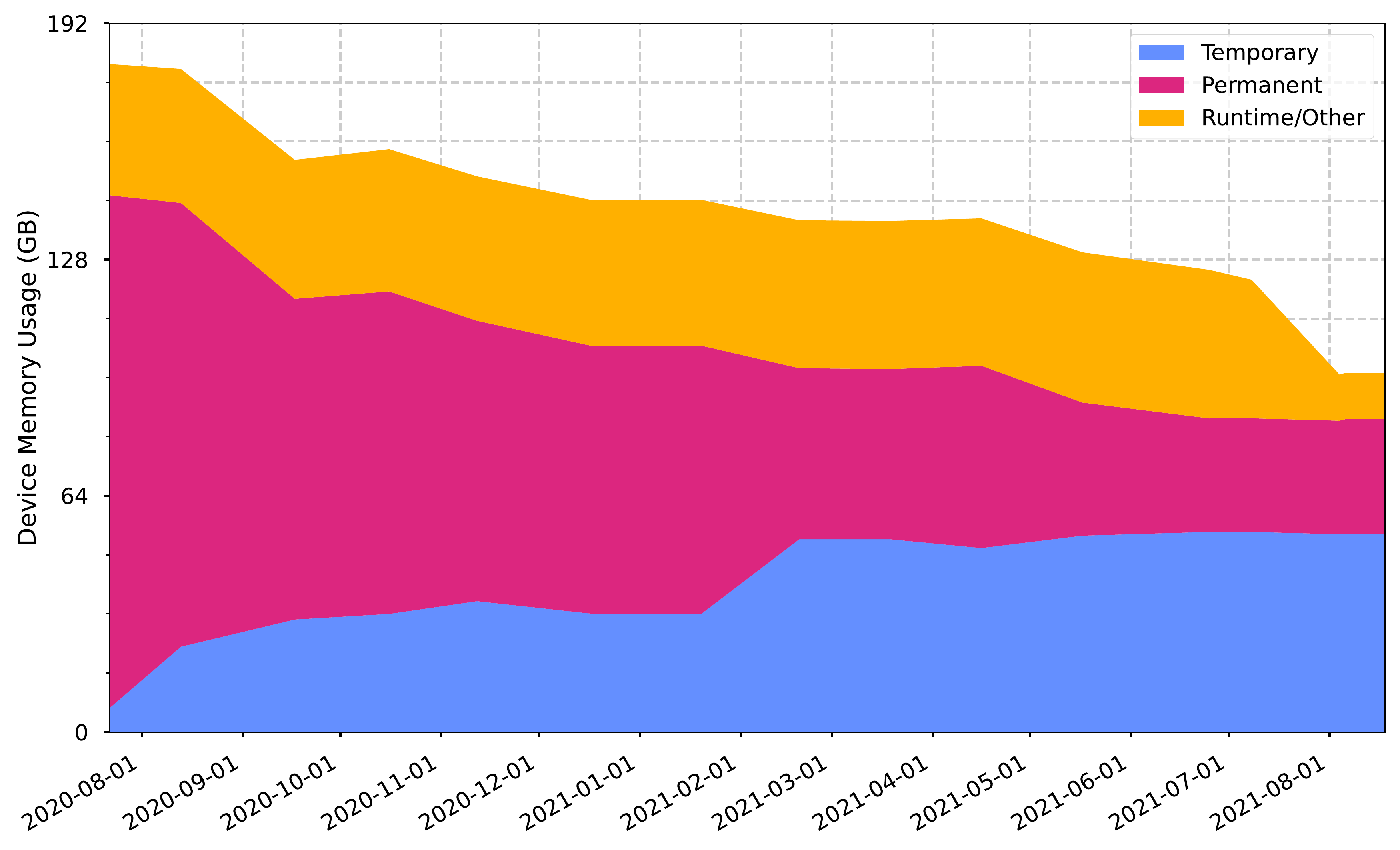}
  \caption{Device memory usage improvements during our GPU port for a 12 GPU 3D Shaped Charge
    \marbl\ simulation with ALE+TMOP. Each compute node has four NVIDIA Volta V100 GPUs with 16~GB of memory.
  }
\label{fig:device_memory_over_time}
\end{figure}

Figure~\ref{fig:device_memory_over_time} shows our overall, temporary,
permanent, and CUDA runtime allocations from when Umpire was introduced to after
we implemented the thread-local-array refactoring.

\section{Cross Platform Scaling Studies}
\label{sec:scaling_studies}

Scaling studies help characterize how well performance scales across large problems
and/or compute resources.
This section describes a set of cross platform scaling studies comparing \marbl's performance across three different
architectures, including a Commodity Technology Systems (CTS) cluster at LLNL, which we refer to as \textsc{CTS-1}
and two Advanced Technology Systems (ATS) -- Sierra, a GPU-based system at LLNL and Astra, an ARM-based system at SNL.
These systems have the following characteristics:

\begin{itemize}[leftmargin=1em]
  \item \textbf{Sierra} is LLNL's 125 petaflop ATS system. It has a Mellanox EDR InfiniBand interconnect
  and 4,320 compute nodes, each of which has two sockets containing 20-core POWER9 processors, 4 NVIDIA Volta V100 GPUs, and 256 GB of memory.
  Our scaling studies used up to 2,048 nodes, comprising half of the machine,
  and ran with 4 MPI ranks per node, with 1 GPU per rank.

  \item  
  \textbf{Astra} is a 2.3 petaflop system deployed under the Sandia Vanguard program.
  It has a Mellanox EDR InfiniBand interconnect and is composed of 2,592 compute nodes, of which 2,520 are user accessible.
  Each node contains two sockets containing 28-core Cavium ThunderX2 64-bit Arm-v8 processors and 128 GB of memory.
  Our scaling studies used up to the full machine (2,496 nodes) and utilized RAJA's OpenMP policy,
  with 2 MPI ranks per node and 28 threads per rank.

  \item
  LLNL's \textbf{CTS-1} clusters are commodity systems with Intel OmniPath interconnect,
  dual 18-core Intel Xeon E5-2695 2.1GHz processors and 128 gigabytes of memory per node.
  Our scaling studies used up to 256 nodes, with 36 MPI ranks per node
  on a system with 1,302 compute nodes and a peak of 3.2 petaflops.

\end{itemize}

Since we are attempting to characterize and compare performance of our code across different computer architectures,
the unit of comparison is an entire compute node of a given architecture.
Such studies, which we refer to as \emph{node-to-node} scaling studies,
differ from typical scaling studies which measure the baseline performance on a single processor.
Node-to-node scaling studies allow natural comparison between a code's performance on different architectures and
its granularity matches how our users typically view their allocations.

Our study includes three types of comparisons:
Strong scaling studies use a fixed \emph{global} problem size and polynomial order and vary the number of compute nodes.
Weak scaling studies use a fixed \emph{local} problem size and polynomial order per compute node and vary the number of compute nodes.
Throughput scaling studies use a fixed compute resource (e.g.\ 1 compute node) and vary the problem size and polynomial order.

Strong scaling demonstrates the ability of a code to solve a fixed problem faster as one adds additional compute resources,
while weak scaling demonstrates the ability of a code to solve a larger problem faster as one adds more resources.
%
%
%
In contrast, throughput scaling characterizes the machine performance and demonstrates the balance between parallelism
and memory capacity; that is, it measures the work required to saturate memory bandwidth on the compute resources.
Commodity systems have modest memory bandwidth and compute resources. As such, they can only accommodate a relatively small
amount of parallelism. They are easy to saturate, yielding overall lower performance.
On the other hand, GPU-based systems, like Sierra, have very high memory bandwidth and compute resources.
As such, they require large amounts of parallel work to saturate, but can yield very high performance.

\subsection{Strong Scaling}
\label{sec:strong_scaling}

  \begin{figure}[tbp]
  \centering
        \includegraphics[scale=0.35]{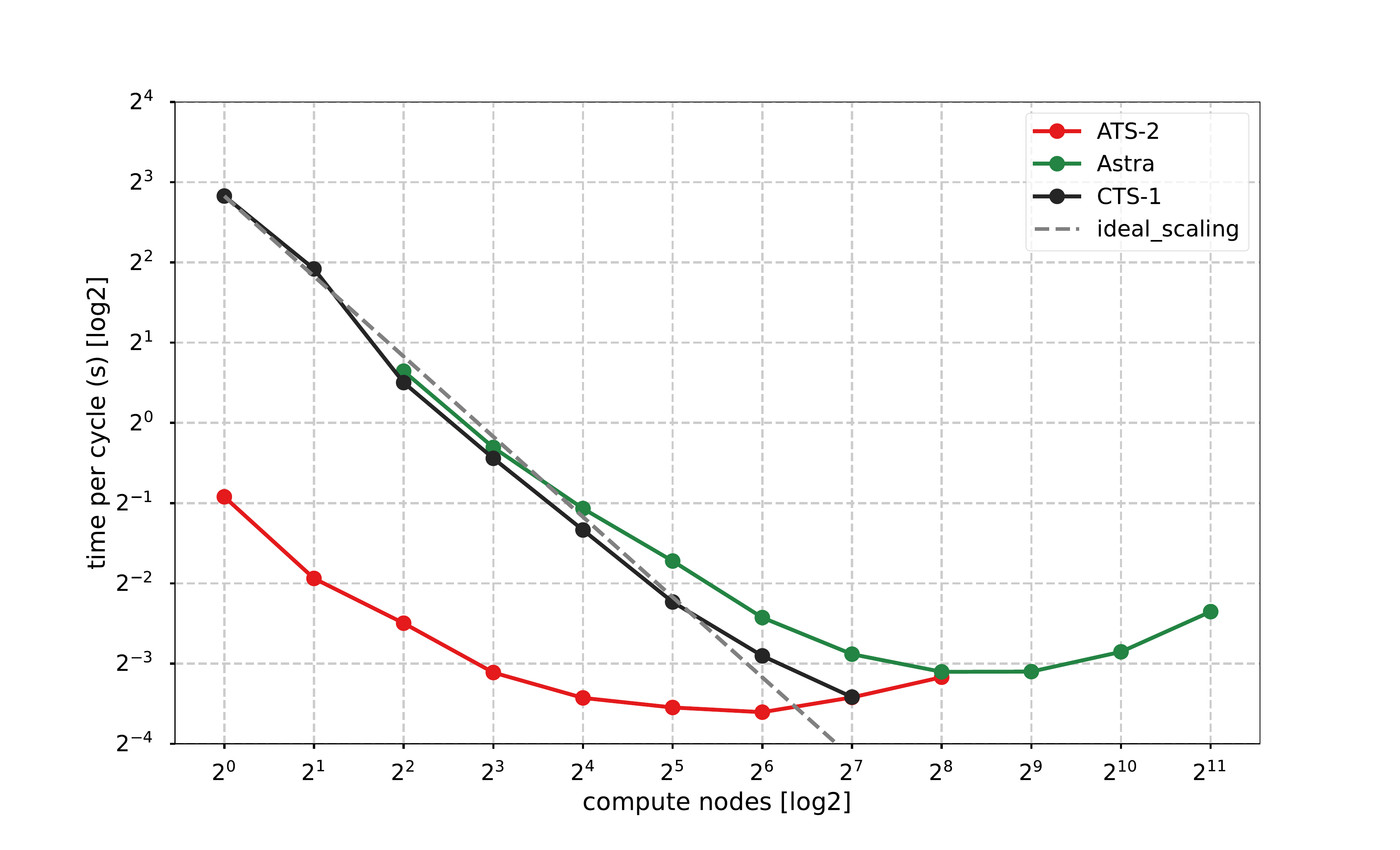}
    \caption{Node-to-node strong scaling for \marbl's Triple-Pt 3D problem with two levels of uniform refinement and TMOP.
      Each data point represents a 500 cycle Lagrange hydro simulation, with an ALE remap every 50 cycles.
      Data is plotted on a log2-log2 scale where the independent axis measure the number of compute nodes
      while the dependent axis measures the time per cycle.
      We also plot ideal strong scaling performance for the CTS-1 series: a line with slope $-1$.}
  \label{fig_marbl_strong_scaling}
  \end{figure}

  Our strong scaling study ran the 3D triple-point shock interaction (Triple-Pt) benchmark, see for example \cite{dobrev2012high}, in ALE mode with TMOP
  with a total of 112,896 quadratic elements and 7,225,344 quadrature points distributed among
  varying numbers of compute nodes.
  Our study starts with one compute node and iteratively doubles the node count in successive runs.
  For ideal strong scaling, the time per cycle would decrease linearly as we increase the node count.
  Specifically, as we double the node count, the time per cycle would reduce by a factor of two.

  Figure~\ref{fig_marbl_strong_scaling} depicts the code's node-to-node strong scaling performance
  on Sierra (red), Astra (green) and CTS-1 (black).
  As can be seen in the figure, \marbl\ exhibits very good strong scaling on the CPU-based architectures
  up to 128 CTS-1 and Astra nodes, when communication costs begin to dominate the runtime.
  As expected, the code exhibits worse strong scaling for higher node counts on the GPU, where,
  as the node count increases, there is no longer sufficient work to feed each GPU.
  As such, for this problem, the strong scaling begins to level off when using
  more than 32 nodes (128 GPUs).

\subsection{Weak Scaling}
\label{sec:weak_scaling}

  \begin{figure}[tbp]
  \centering
    \includegraphics[scale=0.35]{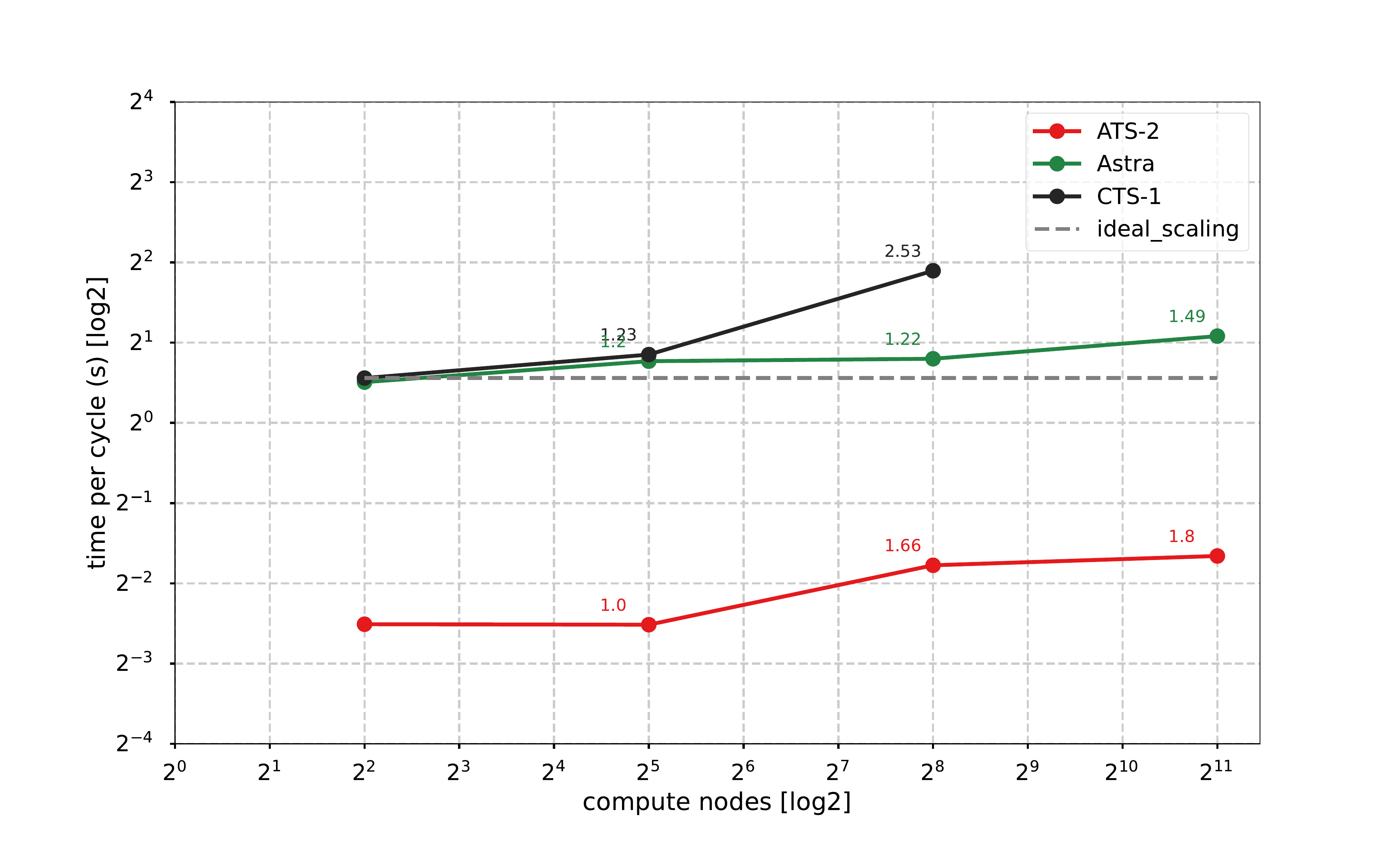}
    \caption{Node-to-node weak scaling for \marbl's Triple-Pt 3D problem on 4, 32, 256 and 2048 compute nodes.
      Each data point represents a 500 cycle Lagrange hydro simulation with an ALE remap step every 50 cycles.
      Data is plotted on a log2-log2 scale where the independent axis is the number of compute nodes
      and the dependent axis is the time per cycle (in seconds).
      The annotations indicate the weak scaling efficiency with respect to the 4 node run
      on that platform.
      Ideal weak scaling would correspond to an efficiency of 1.0 and all data points in that series would lie on a horizontal line.
      }
  \label{fig_marbl_weak_scaling}
  \end{figure}

  Our weak scaling study was performed on the same Triple-Pt 3D problem as Section~\ref{sec:strong_scaling}
  starting with 2 levels of uniform refinement for our first data point.
  Each subsequent run in the series maintained the same average number of
  quadrature points per rank by applying an extra level of uniform refinement (multiplying the global
  problem size by a factor of eight and increasing the node count by a factor of 8).
  As such, we have up to four runs in each of our weak scaling series, using 4, 32, 256 and 2048 nodes.
  We note that 2048 nodes is half of Sierra and 80\% of Astra.

  Figure~\ref{fig_marbl_weak_scaling} depicts the code's weak scaling performance
  on Sierra (red), Astra (green) and CTS-1 (black). Since each rank has the same local problem size,
  ideal weak scaling would maintain the same compute time as we scale out to more nodes.
  This is not generally achievable due to the increased inter-node communication as we scale out to more nodes.
  In addition to plotting the compute time per cycle, we also annotate each data point
  with the weak scaling \emph{efficiency}, measured as the time per cycle of the current data point
  divided by that of the first data point in the series.

  As can be seen in the figure, our code achieves good weak scaling performance as we scale out
  to the maximum number of nodes in the study.
  Comparing Sierra performance to the other platforms, we observe 8--16$\times$ node-to-node speedup
  for Sierra runs compared to those for the CPU-based platforms.

\subsection{Throughput Scaling}
\label{sec:throughput_scaling}

\begin{figure}[tbp]
\centering
  \includegraphics[scale=0.35]{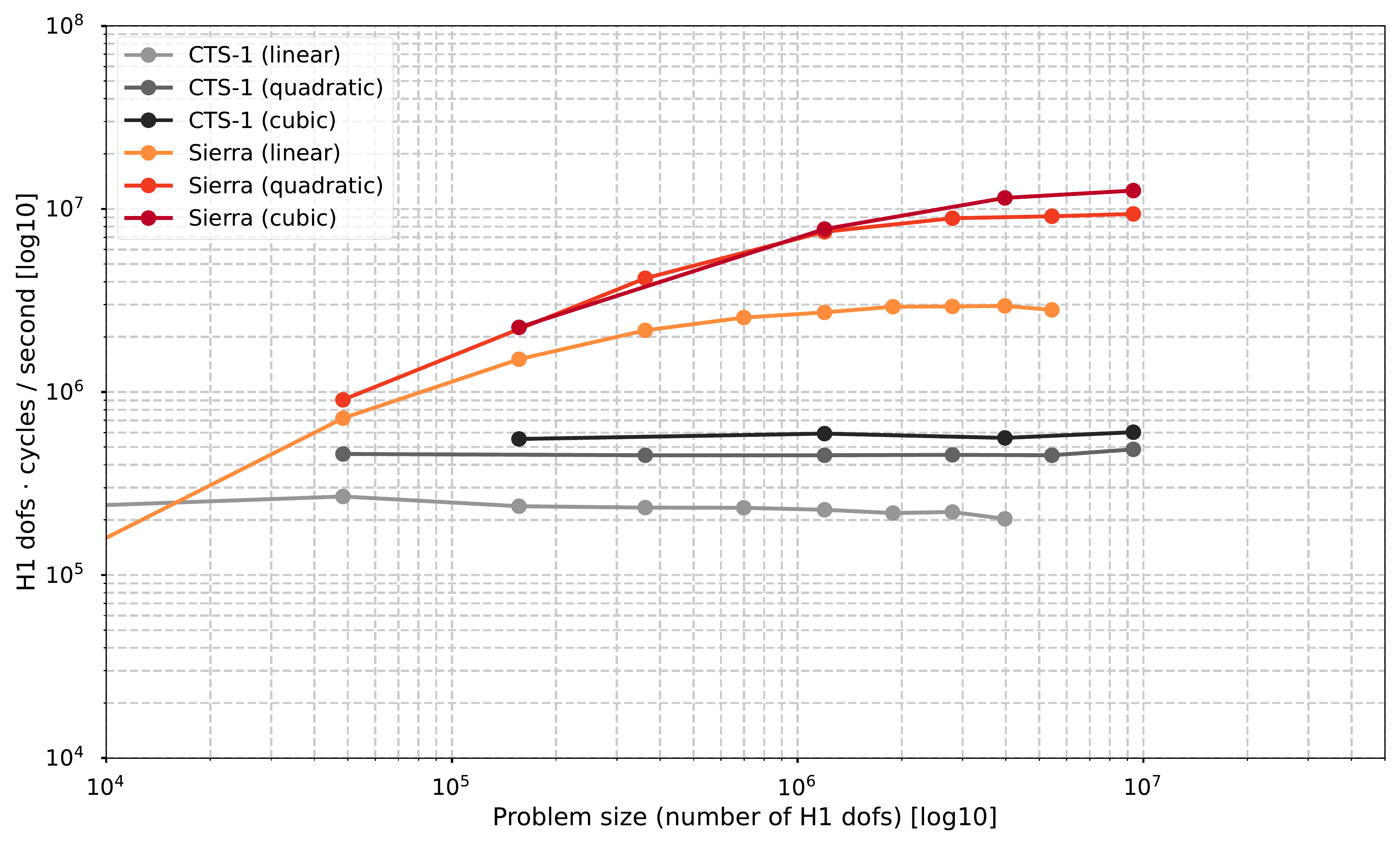}
  \caption{Node to node throughput study for \marbl's Triple-Pt 3D problem
    for varying problem size and polynomial order on a single compute node of Sierra (red series) and CTS-1 (gray series).
    Each data point represents a 500 cycle Lagrange hydro simulation with an ALE remap step every 50 cycles.
  }
\label{fig_throughput_timesteploop}
\end{figure}

\begin{figure}[tbhp]
  \captionsetup[subfigure]{justification=centering}
  \centering
  \subfloat[Node-to-node throughput: Lagrange phase]{
    \includegraphics[scale=0.3]{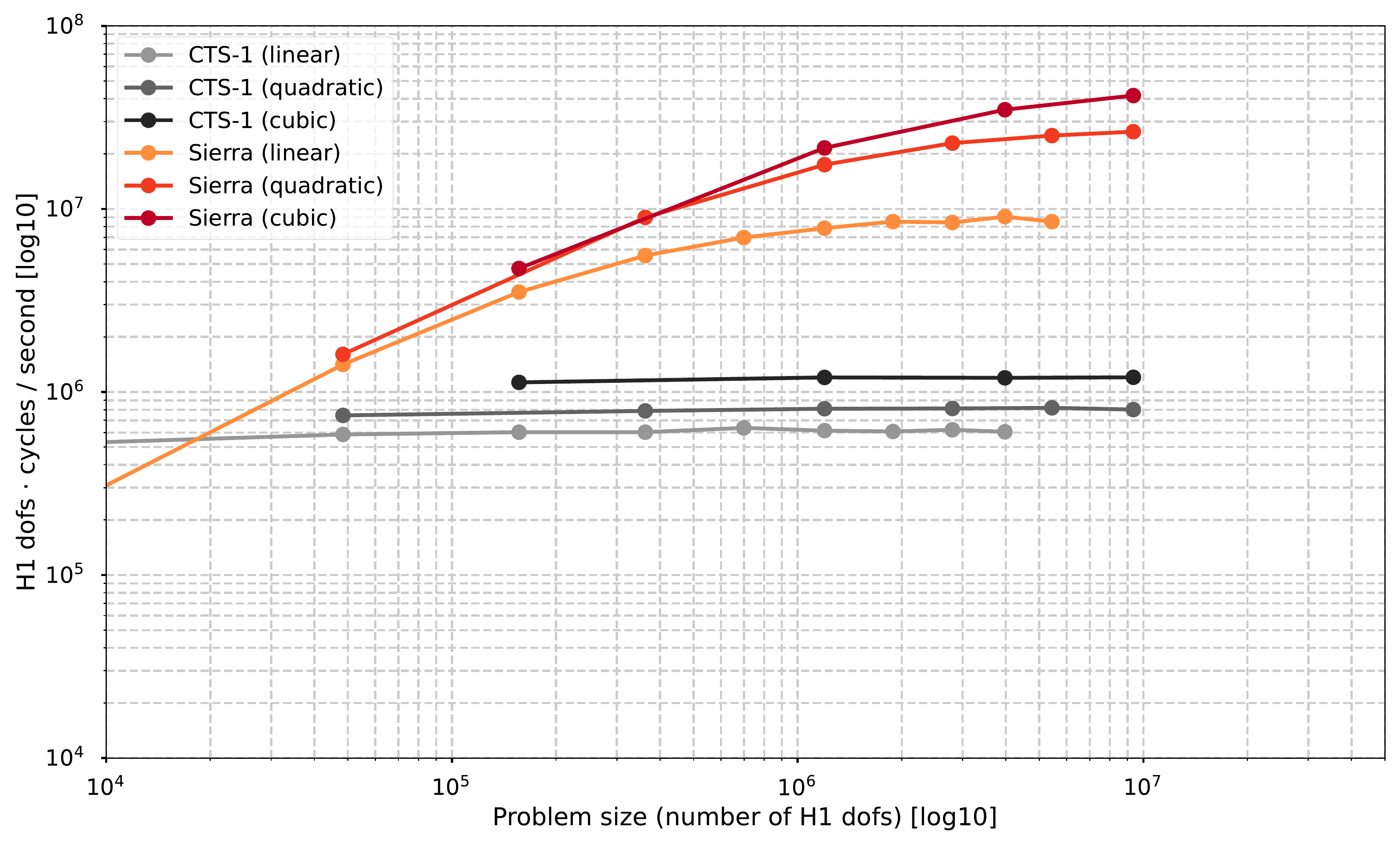}
    \label{fig_throughput_lagrange}
  }
  \\
  \subfloat[Node-to-node throughput: Mesh Optimization phase]{
    \includegraphics[scale=0.3]{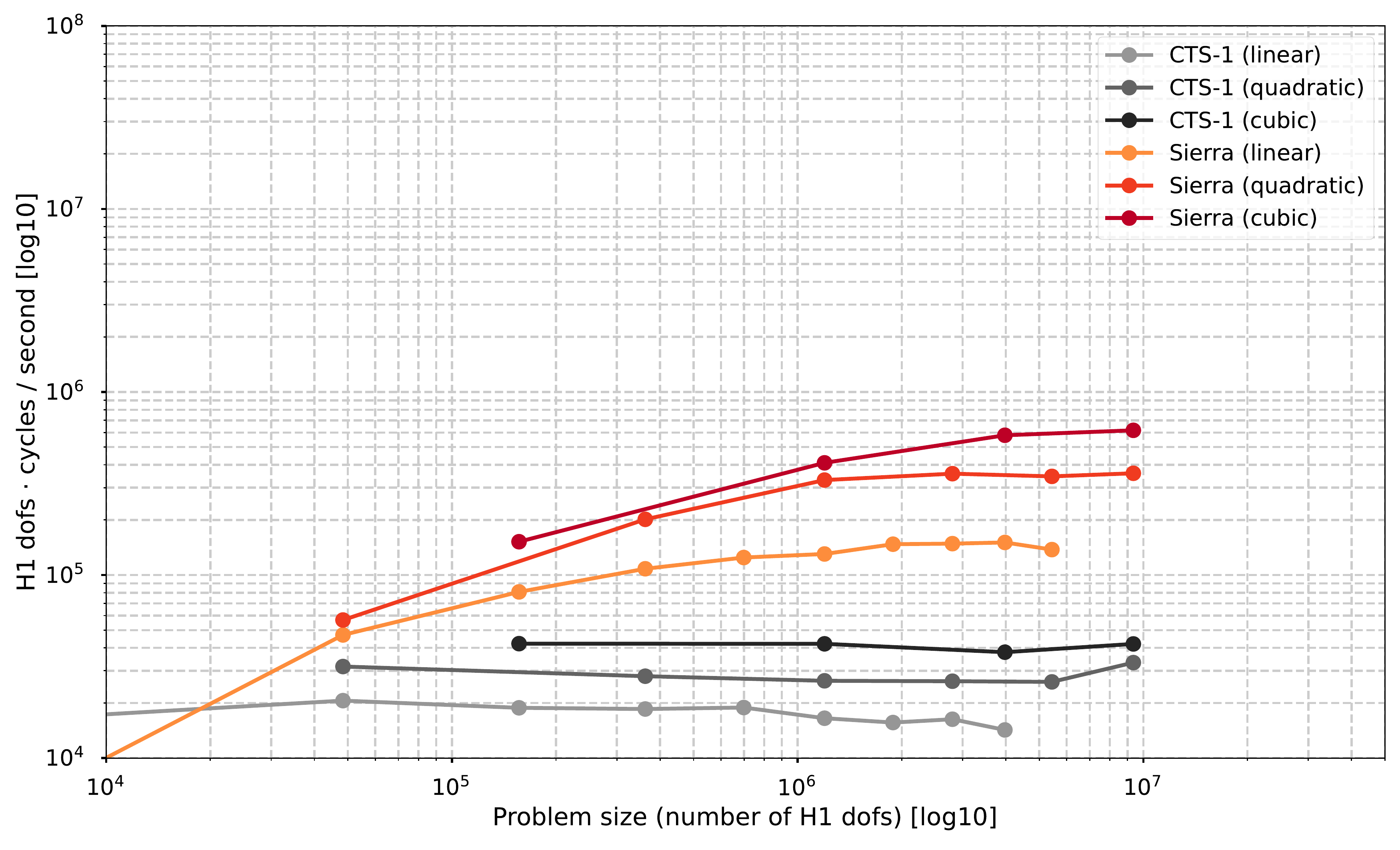}
    \label{fig_throughput_remesh}
  }
  \\
  \subfloat[Node-to-node throughput:  Remap phase]{
    \includegraphics[scale=0.3]{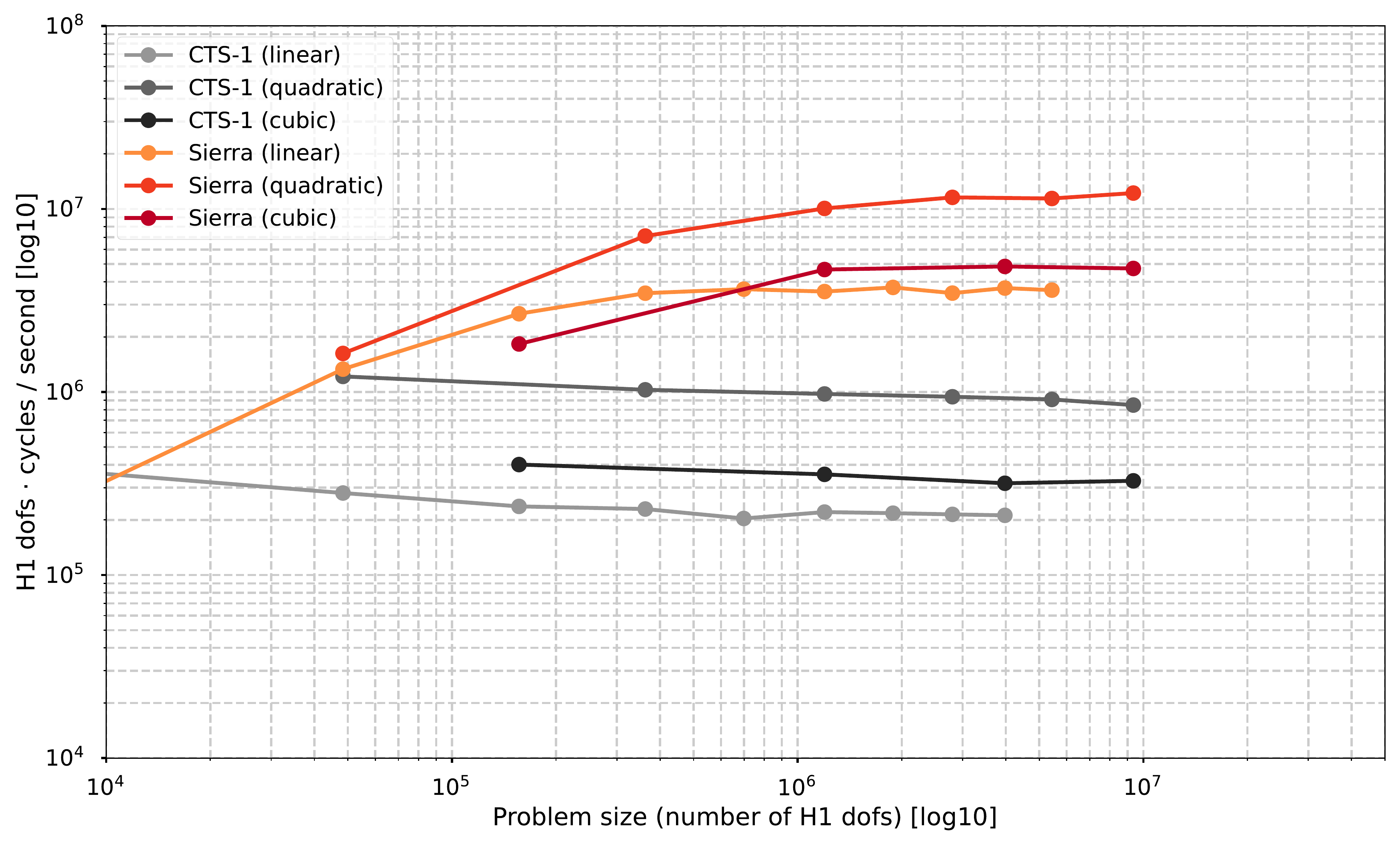}
    \label{fig_throughput_remap}
  }
  \caption{Break-down of the node-to-node throughput study results from Figure~\ref{fig_throughput_timesteploop}
    into its three phases: (a) Lagrange (b) Mesh optimization (TMOP) and (c) Remap.
  }
\end{figure}

\begin{figure}[tbhp]
\centering
  \includegraphics[scale=0.3]{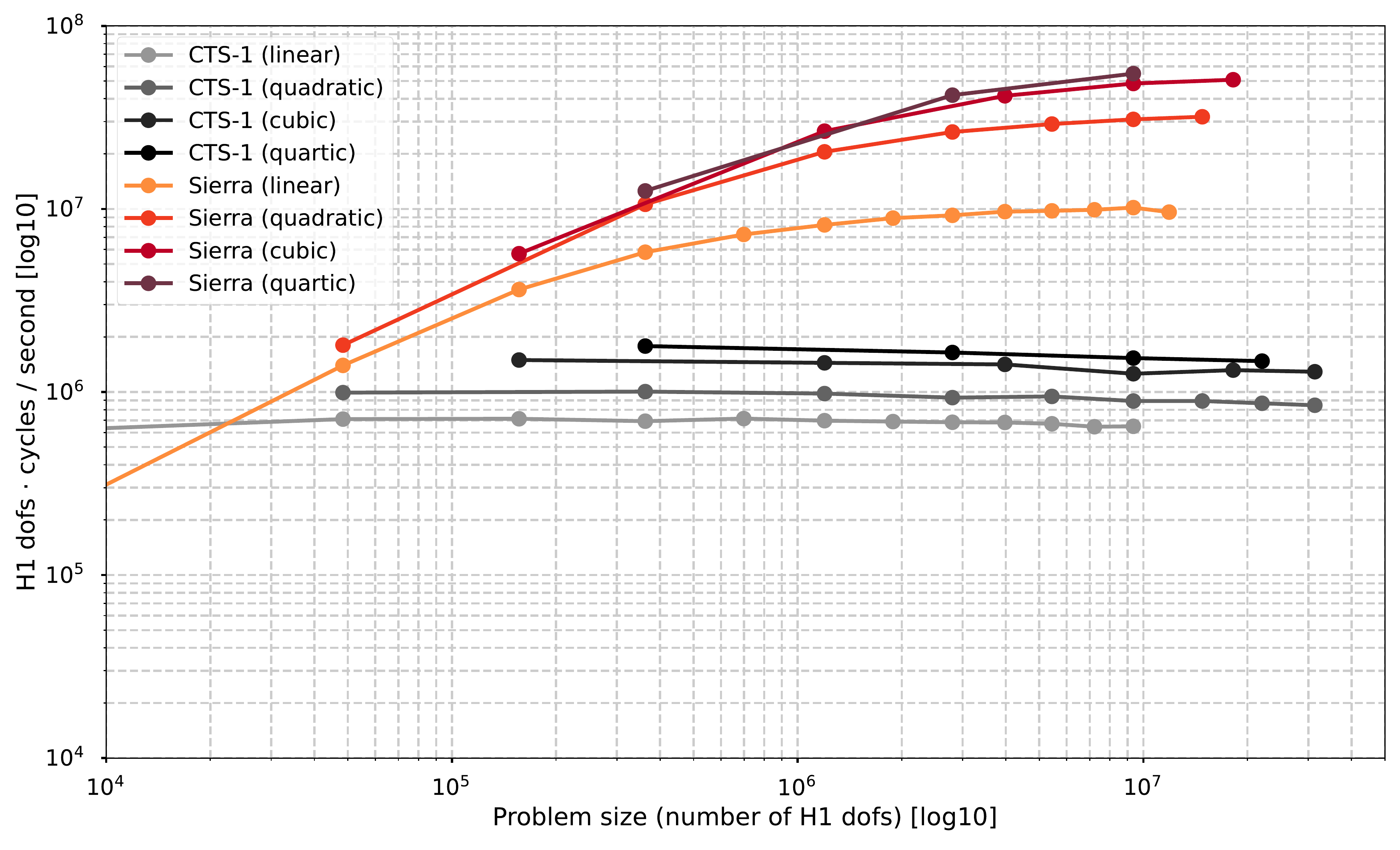}
  \caption{Lagrange throughput study on Sierra for linear, quadratic, cubic and quartic runs.}
\label{fig_throughput_lagrange_quartic_cuda}
\end{figure}

Throughput studies compare the rate an application can process units of work over time
at varying problem sizes. In this section we present throughput studies for linear, quadratic,
and cubic elements by considering the total number of degrees of freedom
for the kinematic variables (H1 basis functions). As can be seen in Figure~\ref{fig_throughput_timesteploop},
\marbl's throughput is significantly higher on GPU-based platforms than on CPU-based ones.
Further, throughput increases both with problem size and with increasing order.
This is in agreement with results shown for Finite Element benchmark problems
in~\cite{ceed-ms6} by the Center for Efficient Exascale Discretizations (CEED).


Breaking out the throughput into the three large phases of \marbl's ALE hydrodynamics
reveals that both the Lagrange phase (Figure~\ref{fig_throughput_lagrange}) and mesh optimization
phase (Figure~\ref{fig_throughput_remesh}) achieve even better throughput with
increasing order.  As can be seen in Figure~\ref{fig_throughput_lagrange_quartic_cuda},
which is restricted to the Lagrange phase on Sierra,
this trend continues for quartic runs.
The remap phase in Figure~\ref{fig_throughput_remap} sees a
decrease in cubic throughput when compared to quadratic but that is expected in our current implementation
which requires full assembly. We are currently investigating a matrix-free remap implementation,
to improve our throughput for cubic and higher orders.

\section{Mini-App Driven Development}
\label{sec_miniapp}

Because \marbl\ was originally developed as a CPU-only code based on full matrix assembly,
the introduction of matrix-free methods required significant restructuring and
rewriting of its major components. To devise a strategy to refactor \marbl,
we developed an MFEM-based miniapp~\cite{laghos}, which serves as a proxy application of \marbl's Lagrangian phase.
Laghos builds on the approach described in \cite{dobrev2012high} modeling the
principle computational kernels of explicit time-dependent shock-capturing
compressible flow in a simplified setting, without the additional
complication of physics-specific code.
Laghos follows a simplified Lagrangian phase of \marbl\ which consist of four major
computations, namely, inversion of a global CG mass matrix, computation of physics and material
specific quadrature point data, computation of artificial viscosity
coefficients, and application of a force operator.

\begin{figure}[tbp]
\centering
  \includegraphics[scale=0.9]{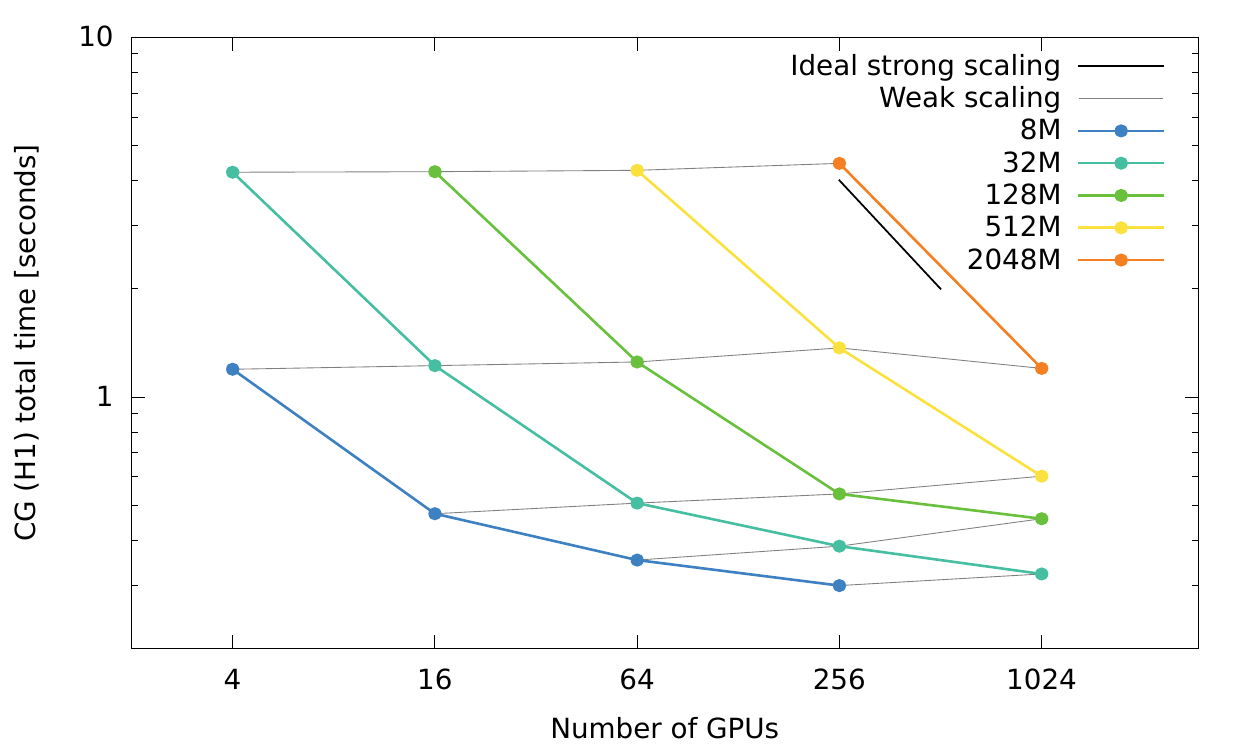}
  \caption{Weak and strong scaling results with Laghos and MFEM-4.1:
           2D problem on Lassen, using up to 1024 GPUs.}
\label{fig_laghos}
\end{figure}

As an initial step for Laghos we derived a code structure that separates the
finite element assembly operations from the quadrature point-based computations.
Such separation is important in the context of matrix-free methods, as it allows
the operator splitting that was discussed in Section~\ref{sec:opDec}.
Based on the new code structure, Laghos was used to introduce the first implementation of
\marbl's partially assembled force and mass operators that produced equivalent
results as the fully assembled versions, but with improved complexity,
see Table~\ref{table:FAvsPA}.
Once the matrix-free CPU algorithms were in place, support was added for
hardware devices, such as GPUs, and programming models, such as CUDA, OCCA, RAJA
and OpenMP, based on MFEM 4.0. Having these different options allowed to perform detailed comparisons of
performance and scaling, e.g., see Figure~\ref{fig_laghos},
and select the most appropriate options for \marbl, based on the available
hardware and the application's preferences.

The work in the Laghos proxy app was of great benefit to \marbl, as it provided
the foundational structure of the code and a baseline version of the major kernels.
Consequently, Laghos served as a model for designing more complex kernels for
\marbl. A key benefit provided to \marbl\ from Laghos and MFEM is the ability to drop in
replacement kernels for standard finite element operations, for example, the action
of the mass operator and diffusion integrator kernels as used in the artificial
viscosity calculations \cite{maldonado2020}.

\section{High-Performance Applications}
\label{sec:high_perf_apps}

We now consider two \marbl\ examples which illustrate the effectiveness of the algorithm
and code improvements previously discussed in the context of 3D high-order multi-material ALE simulations
on GPUs. In each example, we exercise all three phases of ALE using partial assembly techniques
on the GPU including:

   \begin{enumerate}[leftmargin=1em]
    \item High-order Lagrange phase multi-material hydrodynamics on a moving,
       unstructured, high-order (NURBS) mesh~\cite{dobrev2012high},
       including use of GPU accelerated 3rd party libraries for material
       equations of state (EOS) evaluation and material constitutive models

    \item Non-linear, material adaptive, high-order mesh optimization using the
       TMOP method

    \item High-order continuous (kinematic) and discontinuous (thermodynamic)
       Galerkin based remap using flux corrected transport (FCT)
   \end{enumerate}

\subsection{Idealized ICF Implosion}

\begin{figure}[tbp]
\centering
  \includegraphics[height = 0.25\textwidth]{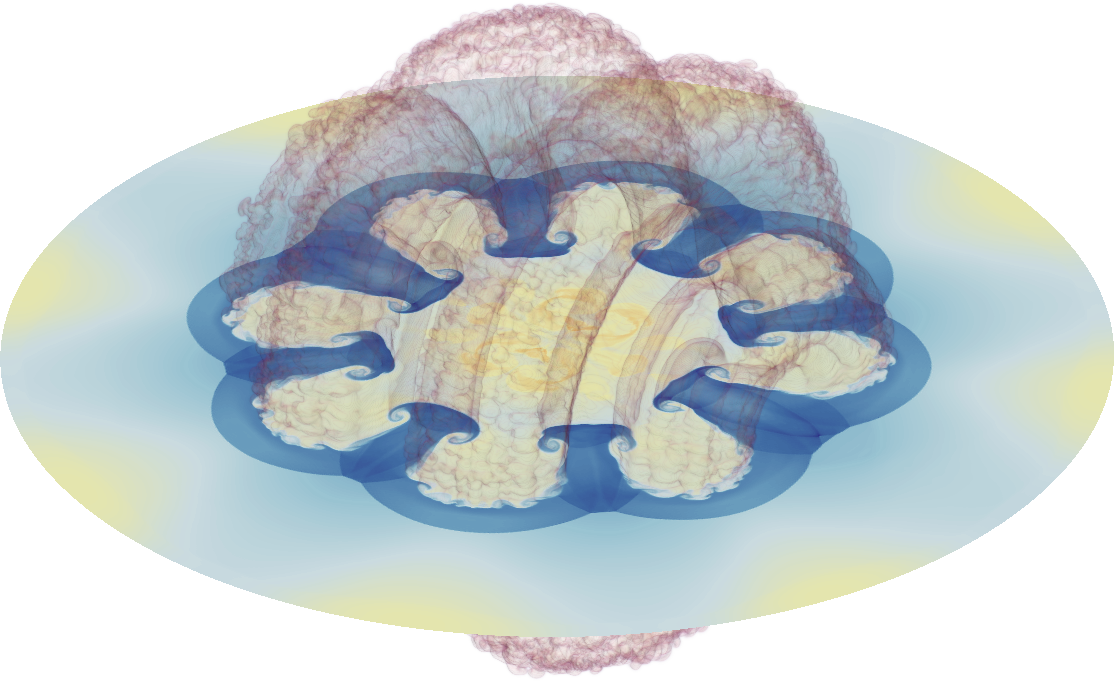}
  \includegraphics[height = 0.25\textwidth]{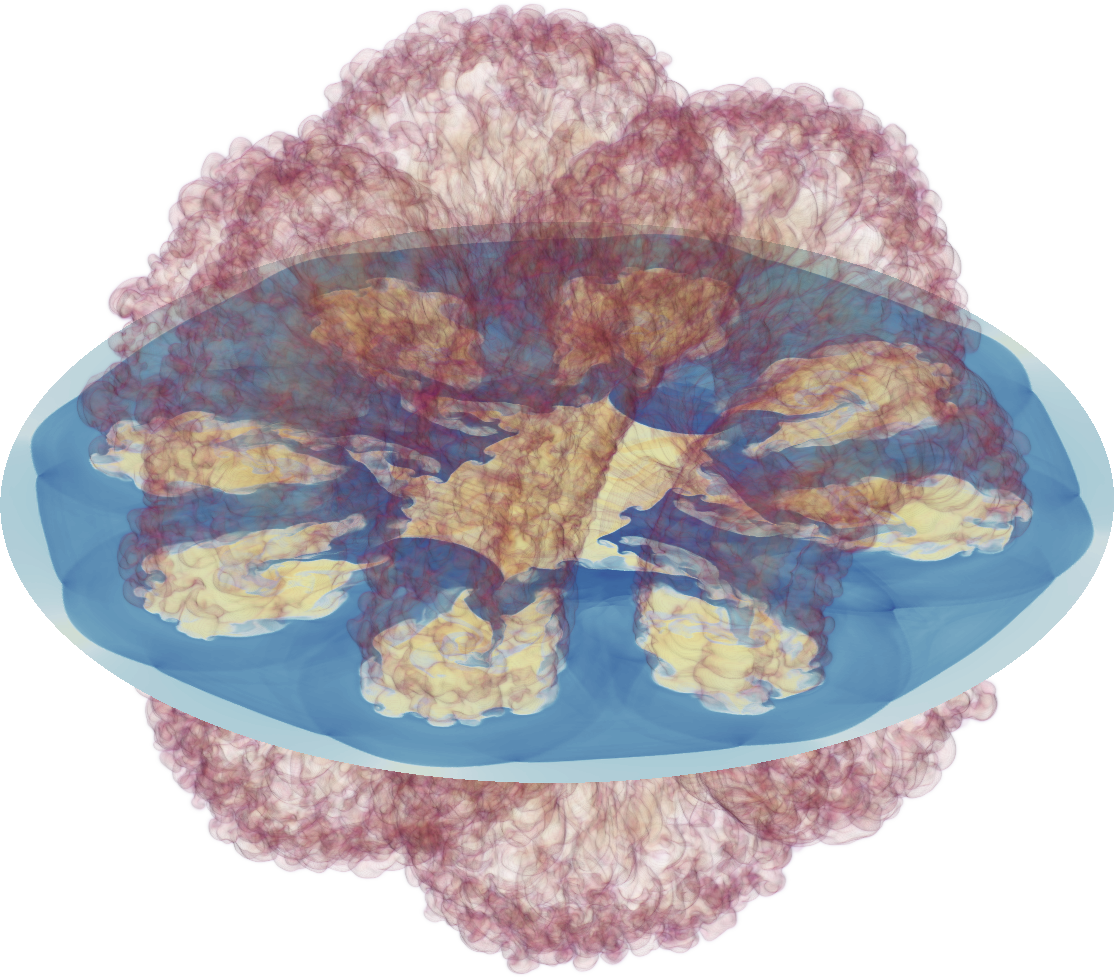}
  \caption{Material density and volume rendering of material interface at two different temporal snapshots for the 3D idealized ICF implosion test
  run on 256 GPUs of the Lassen machine.}

\label{fig_simpleICF}
\end{figure}

\begin{figure}[tbp]
\centering
  \includegraphics[height = 0.25\textwidth]{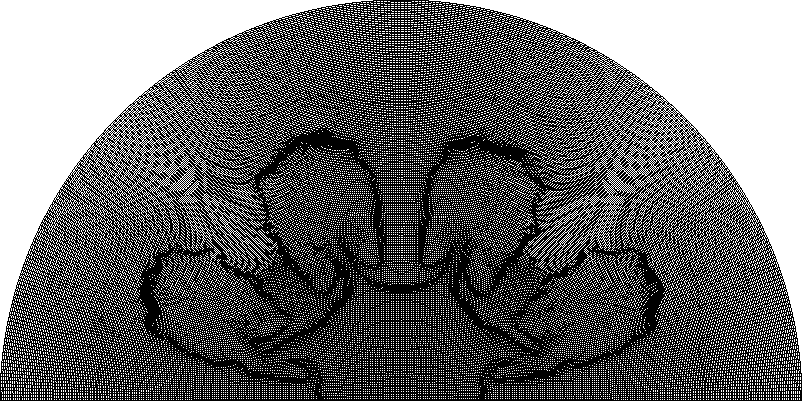}
  \includegraphics[height = 0.25\textwidth]{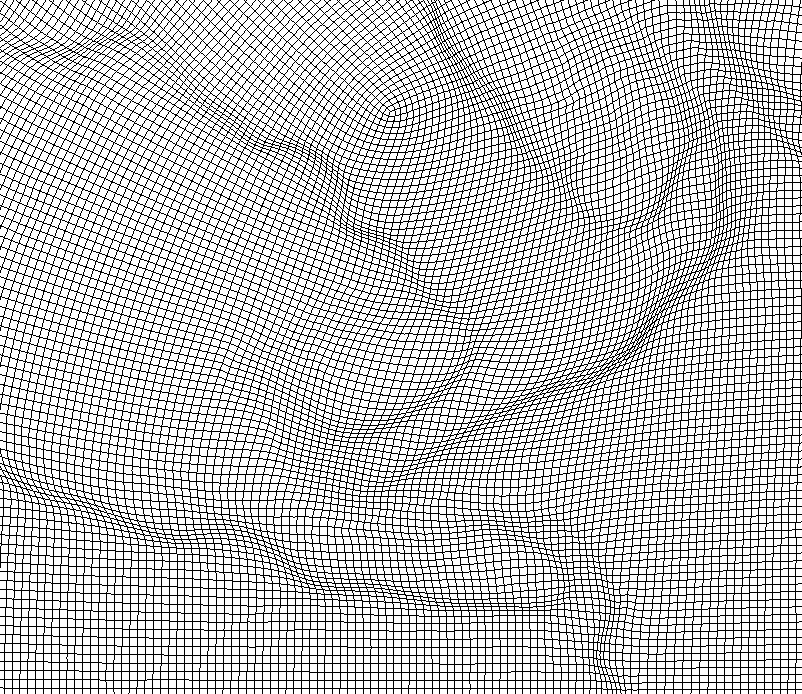}
  \caption{Slice of computational mesh (\emph{left}) with zoomed in view (\emph{right}) highlighting the benefit of material interface adaptive TMOP mesh optimization.}
\label{fig_simpleICFmesh}
\end{figure}

Here we consider a 3D two material idealized ICF implosion test problem which is based on the simplified ICF
benchmark originally described in~\cite{GaleraMaireBreil10}.
In this variant, we consider a 1/4 symmetry 3D simulation domain using a NURBS mesh and apply a constant velocity drive to the outer radius of
the problem (as described in~\cite{DobrevEllisKolevRieben12}). In addition, to generate vorticity during
the implosion, we apply a single mode cosine wave perturbation to the material interface with amplitude $A = 0.02$ and
wavelength $\omega = 8\pi$, specifically we modify the material interface radius as
$r_1(\theta) \mapsto r_1(\theta) + A\cos(\omega \theta)$ as described in~\cite{maldonado2020}. This idealized test problem illustrates the impact of both
the Rayleigh-Taylor and Richtmyer-Meshkov fluid instabilities in three dimensions at the interface between the two
materials as shown in Figure~\ref{fig_simpleICF}. This simulation consists of $\sim\!500~M$ quadrature points and was exercised on 256 GPUs (64 nodes, 4 NVIDIA V100's per node) of LLNL's Lassen machine.
To enhance the high-order mesh resolution near the unstable material interface, we employ
the material adaptive capabilities of the TMOP mesh optimization phase at the material interface with a 2:1 size ratio as shown in Figure~\ref{fig_simpleICFmesh}.

\subsection{Shaped Charge}

   The BRL81a shaped charge is a device which
   focuses the pressures of a high explosive onto a metal ``liner'' to
   form a hyper-velocity jet which can be used in many applications, including
   armor penetration, metal cutting and perforation of wells for the oil/gas
   industry~\cite{walters2008brief}. Modeling such a device requires multi-material compressible
   hydrodynamics with general equations of state for the various materials,
   high-explosive burn and elasto-plasticity.

\begin{figure}[tbp]
  \centering
  \includegraphics[width = 0.33\textwidth]{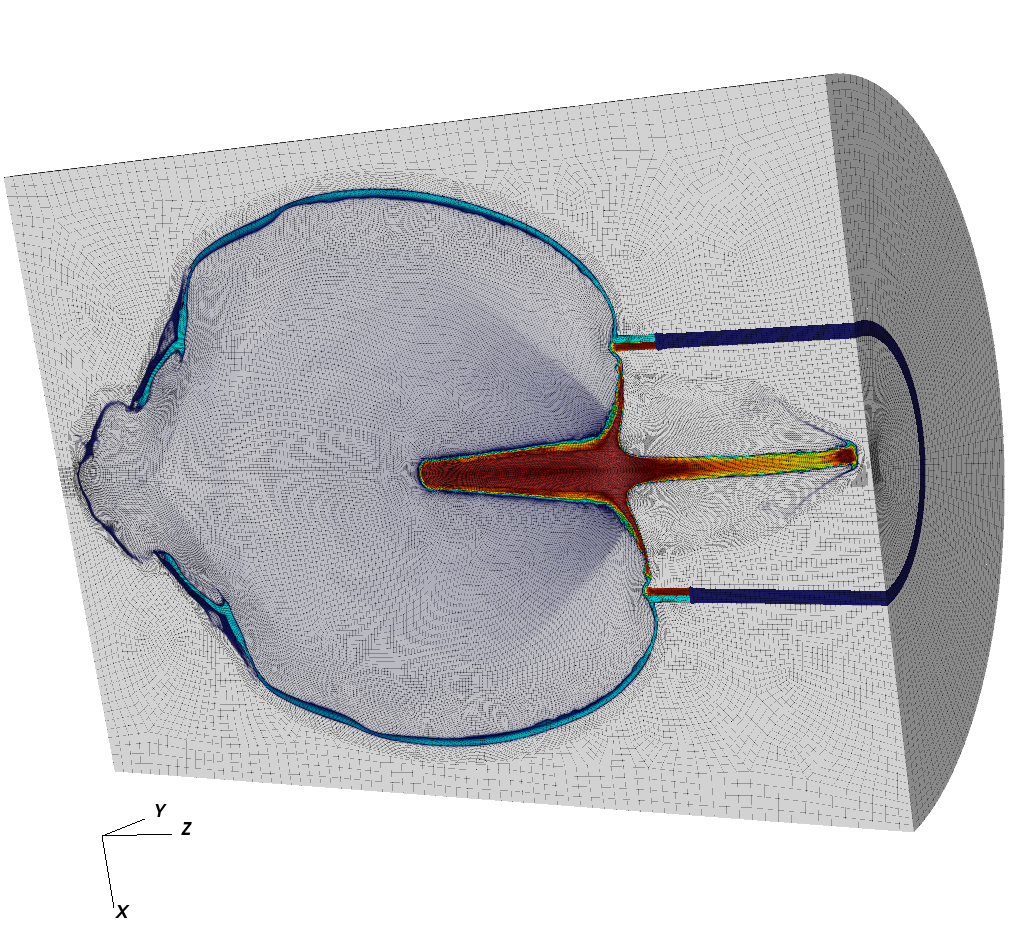}
  \includegraphics[width = 0.33\textwidth]{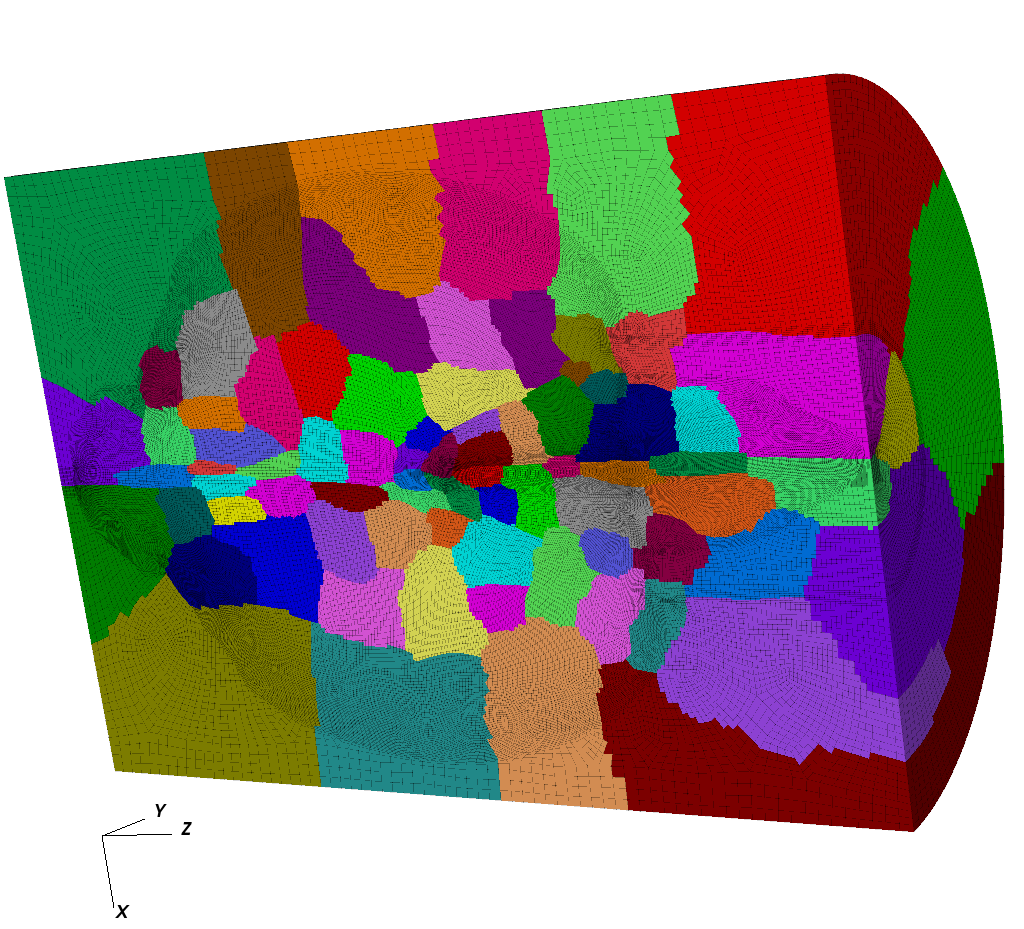}
  \caption{Material density/mesh (\emph{left}) and parallel MPI domains (\emph{right}) for the 3D BRL81a shaped charge
   run on 96 GPUs of the Lassen machine.}
\label{fig_ShapedCharge}
\end{figure}

   In Figure~\ref{fig_ShapedCharge}, we show results of a \marbl\ calculation of the BRL81a with an off-axis high-explosive detonation
   point to illustrate 3D effects. We use the Jones Wilkins Lee (JWL)
   equation of state for the high explosive, tabular equations of state (provided by the LLNL LEOS library)
   for the rest of the materials and a combination of
   elastic-perfectly plastic and Steinberg-Guinan strength models for the solid materials in the problem.
   All of the material constitutive models are running on the GPU either directly in \marbl\ or as a GPU accelerated 3$^{rd}$ party library call.
   The material model libraries perform their
   operations in batches of element quadrature points in parallel on the GPU
   during the Lagrange phase.

\begin{figure}[tbp]
\centering
  \includegraphics[height = 0.25\textwidth]{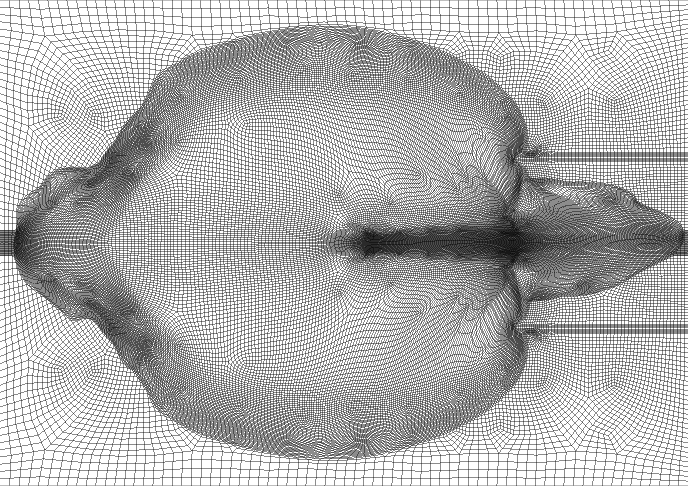}
  \includegraphics[height = 0.25\textwidth]{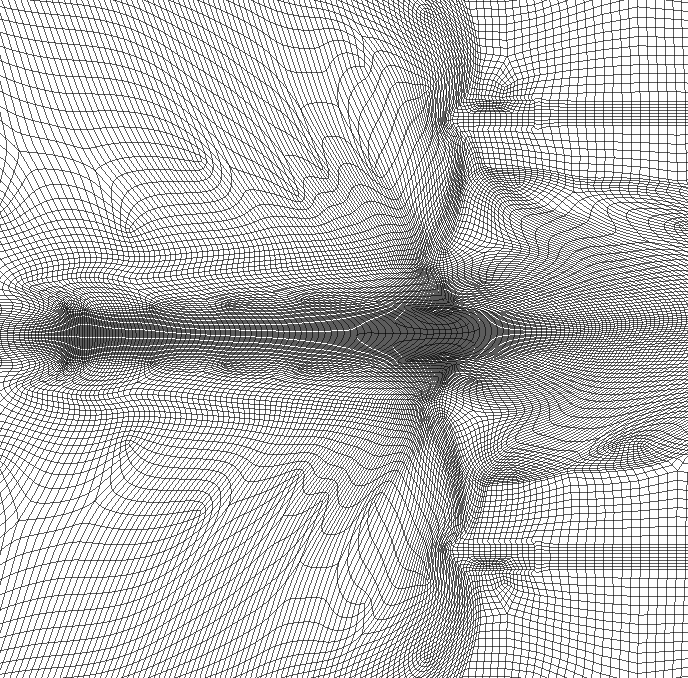}
  \caption{Slice of computational mesh (\emph{left}) with zoomed in view (\emph{right}) highlighting the benefit of material  adaptive TMOP mesh optimization.}
\label{fig_ShapedChargemesh}
\end{figure}

   This problem is a good stress test of the ALE hydrodynamics in \marbl\
   and running this problem at very high resolutions in 3D is important for studying
   the hydrodynamics of hyper-velocity jet formation. There are still
   outstanding questions as to the cause of jet instabilities which are
   experimentally observed. This simulation consists of $\sim\!115~M$ quadrature points and ran on 96 GPUs (24 nodes) of the Lassen machine.
   To enhance the high-order mesh resolution near the hyper-velocity copper jet, we employ
   the material adaptive capabilities of the TMOP mesh optimization phase at the copper material with a 2:1 size ratio as
   shown in Figure~\ref{fig_ShapedChargemesh}.

\section{Conclusion}
\label{sec:conclusion}

In this paper we presented a two pronged approach based on improved algorithms and abstractions to refactoring \marbl, a next-gen code at LLNL, for performance and platform portability across commodity and advanced supercomputer architectures such as the GPU-based Sierra cluster at LLNL and the ARM-based Astra cluster at SNL.
Recognizing changing trends in computing hardware, we began our endeavor by replacing our matrix based algorithms with matrix-free algorithms
wherever possible. The matrix-free algorithms provided improved algorithmic complexity and reduced our memory footprint. To design a single source code which can run on different platforms we integrated the RAJA kernel abstraction layer, Umpire resource memory manager, and MFEM accelerator capabilities. In our efforts we found that extensions and customization of the third-party libraries was crucial for developer productivity and application performance, as was having a proxy app to quickly iterate with different ideas.

To reach improved levels of performance on the \marbl\ code, we co-designed the RAJA Teams API. The RAJA Teams API provided a simplified interface for expressing and reasoning about hierarchical parallelism and portable access to GPU shared memory. Allocating and managing GPU device memory comes with increased complexity due to the high cost of memory allocations and limited GPU memory.
By combining MFEM's memory manager with Umpire, we were able to have MFEM objects quickly draw and release memory from memory pools. The use of memory pools enabled different physics algorithms of \marbl\ to share memory thereby reducing the total amount of permanent memory required across the whole simulation.

We also demonstrated the code's performance on several example problems and with large scale ``node-to-node'' throughput, weak, and strong scaling studies across different supercomputer architectures. Our abstraction model, based on RAJA, Umpire and MFEM, enabled us to run simulations using the same codebase on GPU-based architectures (with MPI+CUDA) as we ran on CPU-based advanced architectures (using MPI+OpenMP) and on commodity platforms using only MPI.

In summary, by incorporating new algorithms and powerful abstraction layers we were able to achieve a high degree of portability and performance with the \marbl\ code.

\section*{Acknowledgments}
 This work performed under the auspices of the U.S. Department of Energy by Lawrence Livermore National Laboratory under Contract DE-AC52-07NA27344. LLNL-JRNL-829593. We would like to thank Veselin Dobrev for support on the integration of MFEM GPU capabilities into MARBL, and the RAJA \& Umpire teams for their feedback in software integration and extensions.
 
\bibliographystyle{unsrt}  
\bibliography{references}

\end{document}